\documentclass[usenatbib]{mn2e}
\usepackage{graphicx,times}
\usepackage{latexsym,amsfonts,amsmath,amssymb}
\usepackage{natbib}
\usepackage{color}
\usepackage[utf8]{inputenc}
\usepackage{ulem}
\usepackage{longtable}
\usepackage{soul}

\title[The merger history of the complex cluster Abell 1758]{The merger history of the complex cluster Abell 1758: a combined weak lensing and spectroscopic view}

\author[Monteiro-Oliveira et al.] 
  {R.~Monteiro-Oliveira,$^1$\thanks{E-mail: rogerionline@gmail.com}
  E. S.~Cypriano,$^1$ R. E. G.~Machado,$^{1,2}$ G. B.~Lima Neto,$^1$
  \newauthor 
  A. L. B.~Ribeiro,$^3$
  L.~Sodr\'e Jr.$^1$ and
  R.~Dupke$^{4,5,6,7}$ \\
  $^1$Instituto de Astronomia, Geof\'isica e Ci\^encias Atmosf\'ericas, Universidade de S\~ao Paulo, R. do Mat\~ao 1226, 05508-090 S\~ao Paulo, Brazil\\
  $^2$Departamento de Ciencias F\'isicas, Universidad Andr\'es Bello, Av.Rep\'ublica 220, Santiago, Chile\\
  $^3$Laborat\'orio de Astrof\'isica Te\'orica e Observacional, Universidade Estadual de Santa Cruz – 45650-000, Ilh\'eus-BA, Brazil\\
  $^4$Observat\'orio Nacional,  Rua Gal. Jos\'e Cristino 77, 20921-400 Rio de Janeiro RJ, Brazil\\
  $^5$Department of Astronomy, University of Michigan, 311 West Hall 1085 South University Ave. Ann Arbor, MI 48109-1107 USA\\
  $^6$Department of Physics and Astronomy, University of Alabama, Box 870324, Tuscaloosa, AL 35487, USA\\
  $^7$Eureka Scientific Inc., 2452 Delmer St. Suite 100, Oakland, CA 94602, USA  
  }

\begin{document}  

\date{Accepted 2016 December 9. Received 2016 November 29; in original form 2016 May 24}

\pagerange{\pageref{firstpage}--\pageref{lastpage}} \pubyear{2016}

\maketitle

\label{firstpage}

\begin{abstract}

We present a weak-lensing and dynamical study of the complex cluster Abell~1758 (A1758, $\bar{z} = 0.278$) supported by hydrodynamical simulations. This cluster is composed of two main structures, called A1758N and A1758S. The Northern structure is composed of A1758NW \& A1758NE, with lensing determined masses of $7.90_{-1.55}^{+1.89}$ $\times 10^{14}$ M$_{\odot}$ and $5.49_{-1.33}^{+1.67}$ $\times 10^{14}$ M$_{\odot}$, respectively. They show a remarkable feature: while in A1758NW there is a spatial agreement among weak lensing mass distribution, intracluster medium and its brightest cluster galaxy (BCG) in A1758NE the X-ray peak is located $96_{-15}^{+14}$ arcsec away from the mass peak and BCG positions. Given the detachment between gas and mass we could use the local surface mass density to estimate an upper limit for the dark matter self-interaction cross section: $\sigma/m<5.83$ cm$^2$ g$^{-1}$. Combining our velocity data with hydrodynamical simulations we have shown that A1758~NW \& NE had their closest approach $0.27$ Gyr ago and their merger axis is $21\pm12$ degrees from the plane of the sky. In the A1758S system we have measured a total mass of $4.96_{-1.19}^{+1.08} \times 10^{14}$~M$_{\odot}$ and, using radial velocity data, we found that the main merger axis is located at $70\pm 4$ degrees from the plane of the sky, therefore closest to the line-of-sight.

\end{abstract}


\section{Introduction}
\label{sec:intro}

Galaxy clusters are currently at the pinnacle of the structure formation process, occupying the top of the mass function of virialized (or quasi) systems in the Universe \citep{ps74}. According to the hierarchical scenario \citep[e.g.][]{lacey93,millennium05,illustris14}, they have been formed through the collapse of overdense regions and posterior growth by the merger of smaller structures \citep[e.g.][for more details]{kravtsov12}. The merging process involves an amount of energy that was not witnessed since the Big Bang \citep{sarazin04}, the reason why merging galaxy clusters are also known as ``astrophysical particle colliders'' \citep{harvey15}.

Unveiling details of the merger process helps understanding the hierarchical growth of the structures in the Universe. Moreover, merging galaxy clusters are excellent astrophysical laboratories for the study of their three main components (sorted here by descending order of mass: dark matter, intra-cluster medium (ICM) and galaxies), as well as for the interaction between them, since each component feels the merger process in a different way.  Galaxies and dark matter interact mainly via gravitational forces whereas the ICM is also subject to hydrodynamical phenomena such as ram pressure \citep{gunn72}, shocks and cold fronts \citep{markevitch_viki07} that slow gas motion sometimes, causing a measurable detachment between this component with respect to the others that are almost non-collisional. This class of systems are known as dissociative mergers \citep{dawson}.

\subsection{Merging clusters as cosmological laboratories}
\label{subsec:merging.as.lab}

The assumption that the matter content of the Universe is composed on its vast majority of
non-interacting cold dark matter (CDM)  has a central role in the $\Lambda$CDM theory of structure formation and evolution \citep[e.g][]{kauffmann93}. Cosmological simulations based on this scenario are able to reproduce very well the observed  matter distribution at large scales \citep[e.g.][]{millennium05,illustris14,alam15}. However on smaller scales there are some inconsistencies, such as the halo core-cusp problem \citep[e.g.][]{deblok10} and 
the  over-prediction  of the amount of substructures \citep[e.g.][]{dubinski91}. These deviations may be due to the effect of astrophysical processes at such small scales. However, a more universal solution for these inconsistencies could be to allow CDM to have a small amount of self-interaction \citep[e.g.][]{spergel00}. With merging galaxy clusters we can estimate the dark matter self-interaction cross section per unit mass
($\sigma/m$) and thus test this hypothesis.

The first weak-lensing studies of the merging cluster 1E 0657-558 \citep[the ``Bullet cluster'',][]{clowe04,clowe06} successfully argued in favour of CDM  against theories of modified gravity \citep[e.g.][]{milgrom83} to explain the observed detachment between the overall mass distribution and the X-ray emitting ICM. In the most likely scenario, favoured by the observations and the modelling of the 3D geometry, the sub-cluster (bullet) passed through the main cluster and this event took place near the plane of the sky. \cite{markevitch04} produced constraints on the CDM scattering depth using three different approaches: the extent of the detachment, the subcluster velocity and the mass loss after core passage. They found $\sigma/m<1$ cm$^2$ g$^{-1}$ as an upper limit for the CDM self-interaction cross section based on their latter approach. Afterwards, \cite{randall08} using numerical simulations found $\sigma/m<1.25$ cm$^2$ g$^{-1}$ as its most reliable estimation based on the absence of a detachment between weak lensing reconstructed mass peaks and galaxy centroids.

The detachment between the baryonic gas and the dark matter in a dissociative  merging cluster could be understood in terms of the scattering depth $\tau$ of the  components: for the gas $\tau \gg 1$, meaning a strong self interaction, whereas for galaxies $\tau=0$, as in practical terms, they only interact with each other gravitationally.  The dark matter is supposed to have an intermediate behaviour, which suggests an upper limit given by:
\begin{equation}
\tau_s=\frac{\sigma}{m}\Sigma_s < 1 \, ,
\label{eq:cross_section}
\end{equation}
which depends on the subclusters' dark matter surface density \citep{markevitch04}.

Only a relatively small number of dissociative merging systems have been discovered so far, mainly due to the fact that the colliding sub-systems spend most of their time at larger distances from each other. Using a sample of 36 merging systems, \cite{harvey15} applied a statistical method to determine $\sigma/m$  based on the ratio of the distances between galaxies and both CDM and ICM distribution  \citep{harvey14},  but their results do not exclude a non interactive CDM with 68 \% c.l.. They also issue a warning that the study of individual systems is indispensable to deal with the particular morphologies and collision geometries. As examples of remarkable merging systems which support the collisionless CDM scenario we can cite MACS~J0025-1222   \citep[the ``Baby Bullet'',][]{babybullet}, A2744 \citep[``the Pandora cluster'',][]{merten}, A2163 \citep[e.g.][]{okabe_bourdin11}, CI 0024+17 \citep{jee07}, A520 \citep[the ``Cosmic Train Wreck'', e.g.][]{a520} and our target Abell 1758, which we will describe in more detail below.

\subsection{Abell 1758}
\label{subsec:A1758}

\begin{figure*}
\begin{center}
\includegraphics[angle=0, width=1.0\textwidth]{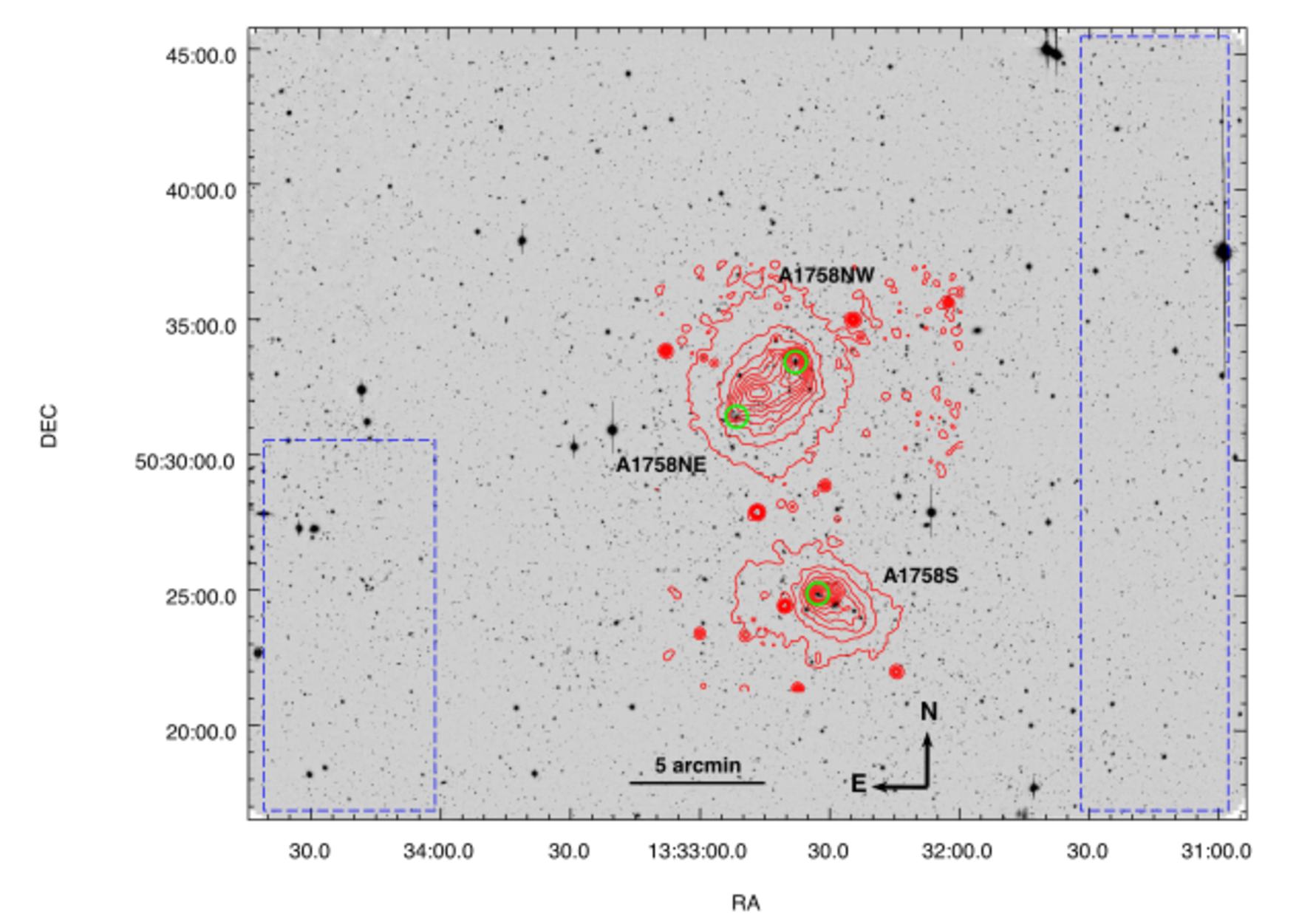} 
\caption{The field of A1758 with the identification of the sub-structures of interest and their respective BCGs (\textit{green} circles). This is a $z'$ band image taken from Subaru Suprime-cam (see Section~\ref{subsec:data_reduction} for a description of the observations) overlaid with Chandra X-ray contours (\textit{red} contours).  The  \textit{blue} dashed rectangles indicate the ``control'' area used for the measurement of the unlensed source density (Section \ref{subsec:kappa_measurement}).}
\label{fig:A1758}
\end{center}
\end{figure*}

First identified by \cite{abell58} as a single richness class 3 cluster, A1758, at $z\sim 0.278$, emerged as a much more complex structure when observed in X-rays by ROSAT \citep{rizza98}. The main cluster structure, which we call A1758N in the present work, appeared as bimodal, with X-ray emission peaks following galaxy overdensities. About 8~arcminutes ($\sim 2$~Mpc) to the south, another concentration of X-ray emitting gas associated with galaxies was found. We call this structure A1758S here. We named the two sub-structures in the North as A1758NW and A1758NE\footnote{\citet{rizza98} named the later as A1758SE but we refrain to use this notation to avoid confusions with the Southernmost structure} (Fig. \ref{fig:A1758}).

Further X-ray observations with Chandra and XMM-\textit{Newton} \citep{davidkempner} reinforced the idea that A1758N\&S are actually
gravitationally bound, but without signs of interaction between them. Their analysis also have favoured A1758N itself as a post-collision system. 
Using density jump measurements, they have estimated the recessional velocity between them as $\sim1600$~km~s$^{-1}$.
They also suggested that A1758S is also a double system, probably at an earlier stage of a merger, which would be happening near the line of sight.

\cite{dahle02} published the first gravitational weak lensing map of A1758, using data from the 2.5~m Nordic optical telescope. The mass distribution of A1758N appeared as a bimodal structure, whereas A1758S appeared as a single clump. \cite{durret11}, using optical (CFHT), X-ray (XMM-\textit{Newton}) and simulated data, obtained galaxy luminosity functions, intracluster gas luminosity and metallicity maps. Their analysis further reinforces the idea of A1758N as a bimodal structure and, mostly, its post-merger status.

Combining weak lensing mass reconstruction made from  Subaru telescope ($R_C$ and $g'$ bands) plus XMM-\textit{Newton} data, \citet{sevenmergers} verified, within the uncertainties, a spatial coincidence in A1758NW among the BCG, the mass and the X-ray peaks. For A1758NE, on the other hand, the BCG position and the corresponding mass peak coincide, but the (rather diffuse) X-rays are detached from both and are located approximately midway between the NW and NE clumps. This configuration resembles that of the ``Bullet cluster'' but with an important difference. For A1758N only one of the components has its gas clearly detached, whereas in the case of the Bullet this occurs for both cluster and sub-cluster. 
\cite{ragozzine} refined the weak-lensing mass maps using Subaru and HST data and determined an overall mass of $2.2\pm 0.5 \times 10^{15}$ M$_{\odot}$ for A1758N.

In a previous paper we \citep{Machado+2015} published hydrodynamical simulations of a system inspired by A1758N. We simulated an off-axis collision of two equal mass ($\sim 5\times 10^{14}$ M$_{\odot}$) clusters and managed to create a intracluster gas morphology similar to that of A1758, where the detachment would be observed in only one of the structures, by allowing different ICM concentrations.

\cite{boschin} performed a dynamical analysis using 92 cluster member redshifts and showed that the bimodality seen in the plane of the sky, in optical or X-ray data, could not be detected in radial velocity space. This fact strongly suggests that most of the motion vector of both spatial sub-systems is mostly perpendicular to our line-of-sight. Those authors also calculated A1758N dynamical mass. This result, among others compiled from the literature, can be seen in Table \ref{tab:mass_comparison}.

\begin{table*}
\begin{center}
\caption[]{A compilation of A1758N mass estimations in the literature}.
\begin{tabular}{l l l r}
\hline \hline 
Mass ($10^{14}$ M$_{\odot}$) $h_{70}^{-1}$ & radius (Mpc)  & method & Reference\\
\hline 
 $16$      	 & $2.6 $                  &  $M \times T_{X}$ scaling relation  &  \cite{davidkempner}	\\	
 $\;\;5.26 \pm 5.70$ & $1.79$  & WL - NFW                              &  \cite{sevenmergers}\\      
 $\;\;4.89\pm1.54$   & $1.3$  & WL - aperture mass densitometry   &  \cite{sevenmergers}\\     
 $22 \pm 5$      & $2.3$                   & WL - aperture mass densitometry   & \cite{ragozzine}\\          
 $2-3$		 & $2.1$                   & Member galaxy dynamics          &  \cite{boschin}\\             
 $12.6 \pm 1.4$  & $1.491 $ 		   & WL - aperture mass densitometry &  \cite{cccp} \\                 
 $11.2 \pm 1.9$  & $1.45$  & WL - NFW fit		 &   \cite{cccp} \\               
 $19.4 \pm 3.2$  & $2.76$  & WL - NFW fit		 &   \cite{cccp} \\  
 $13.39 \pm 1.43$& $2.09$                  & WL - NFW fit                                      & This work (Section~\ref{sec:mass_results})\\
\hline \hline 
\end{tabular}
\label{tab:mass_comparison}
\end{center}
\end{table*}

Even given the wealth of observation of this cluster, there are still open issues that we address with new data, both deep imaging and spectroscopy. In particular we want to quantify objectively the statistical significance of the detachment between gas and mass in A1758NE,  
address the dynamics of this particular merger that produced this  configuration and to estimate upper limits for the dark matter self-interaction cross section. Moreover, A1758S has no optical nor spectroscopic study published yet.

This paper is organised as follows. In Section \ref{sec:wl_analysis} we describe the imaging data and reduction and the weak lensing analysis. In Section~ \ref{sec:dynamical_analysis} we consider the spectroscopic data, its reduction and the dynamical analysis. Our new, tailor made, hydrodynamical simulations for this system, can be found in Section~\ref{sec:simulations}. The proposed merger scenario for A1758N is described in Section~\ref{sec:discussion}, including an estimate of the dark matter self-interaction cross section. In Section \ref{sec:conclusion} we summarize our results on A1758.

Throughout this paper we adopt the following cosmology: $\Omega_m=0.27$, $\Omega_\Lambda=0.73$, $\Omega_k=0$, and $H_0=70$~km~s$^{-1}$ Mpc$^{-1}$.  At mean cluster redshift $z=0.278$ we have $1^{\prime\prime}=4.25$ kpc, an Universe age of $10.6$ Gyr and an angular distance of $876.3$~Mpc.

\section{Weak gravitational lensing analysis}
\label{sec:wl_analysis}

In this section we will describe a novel weak-lensing analysis of this cluster using both distortion and magnification effects with deep three-band Subaru images. 

\subsection{Fundamental Concepts}

As gravitational lensing has become a new standard in astrophysics we refrain to describe all its formalism in here. Instead, we refer the reader to one of the many good reviews such as \citet{mellier99} or \citet{schneider05}, and focus only on the concepts we are going to use more directly.

The gravitational lensing field can be described by a scalar, the convergence ($\kappa$), and a spin-2 tensor, the shear ($\mathbf{\gamma} = \gamma_1 + i\gamma_2$), both second derivatives of the projected gravitational potential. The two components of the shear can be described in the Cartesian framework as a component in the $x-y$ directions (frequently called $\gamma_+$) and one 45 degrees in relation to that ($\gamma_\times$).

The convergence, $\kappa$, is physically the projected surface mass density of the lens in units of the lensing critical density:
\begin{equation}
\Sigma_{cr} = \frac{c^2}{4\pi G} \frac{D_s}{D_{ds} D_d} \, ,
\label{eq:sig_crit}
\end{equation}
where $D_s$, $D_{ds}$ and $D_d$ are, respectively, angular diameter distances to the source, between the lens and the source\footnote{Background objects with respect to the lens, or A1758 in the case.}, and to the lens.

In the weak regime ($\kappa = \Sigma/\Sigma_{cr}\ll 1$) the convergence is associated with an isotropic magnification of the sources. The expected value of the average ellipticity of a sample of unlensed sources is supposed to be zero. The lensing effect, however, coherently distorts those images is a way that the average ellipticity will then tend to the effective shear or distortion ($g$):
\begin{equation}
\langle e \rangle \simeq g \equiv \frac{\mathbf{\gamma}}{1-\kappa} \, ,
\end{equation}
where the ellipticity modulus, in terms of the semi-major ($a$) and semi-minor ($b$) axis, is
\begin{equation}
e = \frac{a-b}{a+b} \, .
\end{equation}

The full ellipticity (as well as the shear and the effective shear) cannot be defined by its modulus only, because it also has an orientation ($\theta$, e.g., the direction of the semi major axis). It is commonly described as a spin-2 tensor (a ``headless'' vector) whose two components can be defined as:
\begin{equation}
e_1 = e \times \cos{(2\theta)}\, ,
\qquad
e_2 = e \times \sin{(2\theta)}\, .
\end{equation}

The distortion of the background population has been measured for several clusters with success for many years now \citep[e.g.][and references within]{kneib_nata11},  but the potential of the magnification effect has begun to be exploited systematically much more recently \citep[e.g.][]{umetsu_broadhurst08,umetsu_broad11}.

The lensing induced magnification suffered by a background source is given by:
\begin{equation}
 \mu=\frac{1}{(1-\kappa)^2- |\gamma|^2}
\label{eq:magnification}
\end{equation}

This magnification can be understood as a stretching of the projected area of the source, whereas its surface brightness is conserved \citep{schneider_ehlers_falco92}. As it increases\footnote{This is always the case in the weak lensing regime, but the same is not true for strong lensing.} the flux of the sources, it stretches the projected space as well, by moving objects radially away from the lens centre. Whereas the former effect tends to increase the observed number density of sources, as we can detect sources that would, otherwise, be below the limiting flux of an observation, the latter has the opposite effect, given that it increases the area which contains the sources.  

The net result on the number density would then go with:

 \begin{equation}
 	 N(<m,r)\approx N_0(<m)\mu(r)^{2.5\alpha-1},
  	\label{eq:bias_mag}
\end{equation}
where $\alpha=d\log N(<m)/dm$ is the logarithmic slope of the galaxy count per magnitude and $N_0(<m)$ is the number density of the unlensed sources.

Therefore, this so called magnification bias behaviour is defined by the slope of the counts: for steeper counts ($\alpha>0.4$) there is an observed increase on the galaxy density due to the lens, whereas the opposite (a decrease) happens when the slope of the counts is shallower ($\alpha<0.4$).

In the weak-lensing regime, where $\kappa, \gamma \ll 1$, one has $g \sim \gamma$ and $\mu\sim 1+2\kappa$, so it is not unusual to associate the observables- image distortions and magnification biases- with the main lensing property involved: shear and the convergence, respectively. In this work we use the observable names given that we also probe regions close to the cluster centres which are not strictly in the weak-lensing regime.

\subsection{Imaging data: observation and reduction}
\label{subsec:data_reduction}

Imaging data were taken with the Suprime-Cam of the Subaru telescope during semester 2007A in queue mode. The observations have been done in the $B$, $R_C$ and $z'$ bands. Further details are summarized in Table~\ref{phot}.

\begin{table}
\begin{center}
\caption{Imaging data characteristics}
\begin{tabular}{lccc}
\hline
\hline
Band & Total exposure (h) & {\it Seeing} (arcsec) & Completeness\footnotemark[1] \\
\hline
$B$     & 2.17 & 1.11 & 26.6 \\ 
$R_C$ & 2.64 & 1.11 & 26.4  \\
$z'$    & 1.00 & 1.02 & 25.7 \\
\hline
\hline
\multicolumn{4}{l}{$^1$ Corresponds to the magnitude in which the logarithmic}\\
\multicolumn{4}{l}{counts drop 5\% by comparison with the Subaru Deep Field }\\
\multicolumn{4}{l}{\citep{sdf}.}
\end{tabular}
\label{phot}
\end{center}
\end{table}

The overall image processing procedure was done using the semi-automatic routine {\sc SDFRED} \citep{sdfred1,sdfred2}, designed for this particular instrument. It consisted of ($i$) bias and overscan subtraction, ($ii$) flat-fielding, ($iii$) atmospheric and dispersion correction, ($iv$) sky subtraction, ($v$) auto guide masking, ($vi$) alignment (done for all three filters simultaneously), ($vii$) combining and mosaicing into the final image per filter, ($viii$) fringing removal from $z'$ images, and ($ix$) registering and combining the images (with {\sc IRAF}).

Most of the observations were taken in photometric time and thus we used standard star fields \citep{landolt92,smith02} to calibrate the magnitudes in the AB system. Some of the $R_C$ images were observed in non-photometric time. To deal with that we used another well calibrated  $R_C$ image from another field we observed within the same program and calibrate a relation between the Subaru $R_C$ and SDSS $g$ and $r$ magnitudes for galaxies, which we later used for A1758.

We built object catalogues with {\sc SExtrator} \citep{sextractor} in ``double image mode'', where the detections were always made in the  $z'$ band, that turned out to be the deepest one. Galaxies were selected among all the objects in the catalogue according to two complementary criteria: for $18.5 \leq R_C \leq 25.75$ galaxies were the objects with FWHM $>$ 1.03 arcsec and for the brightest objects ($R_C<19$), galaxies were identified as having SExtrator's CLASS\_STAR $<$ 0.8. 
Stars (actually point sources) were selected by its stellarity index (bright saturated stars) or its FWHM.

\subsection{Distortion}

\subsubsection{Source selection} 
\label{subsub:background_selection}

\begin{figure}
\begin{center}
\includegraphics[angle=0, width=\columnwidth]{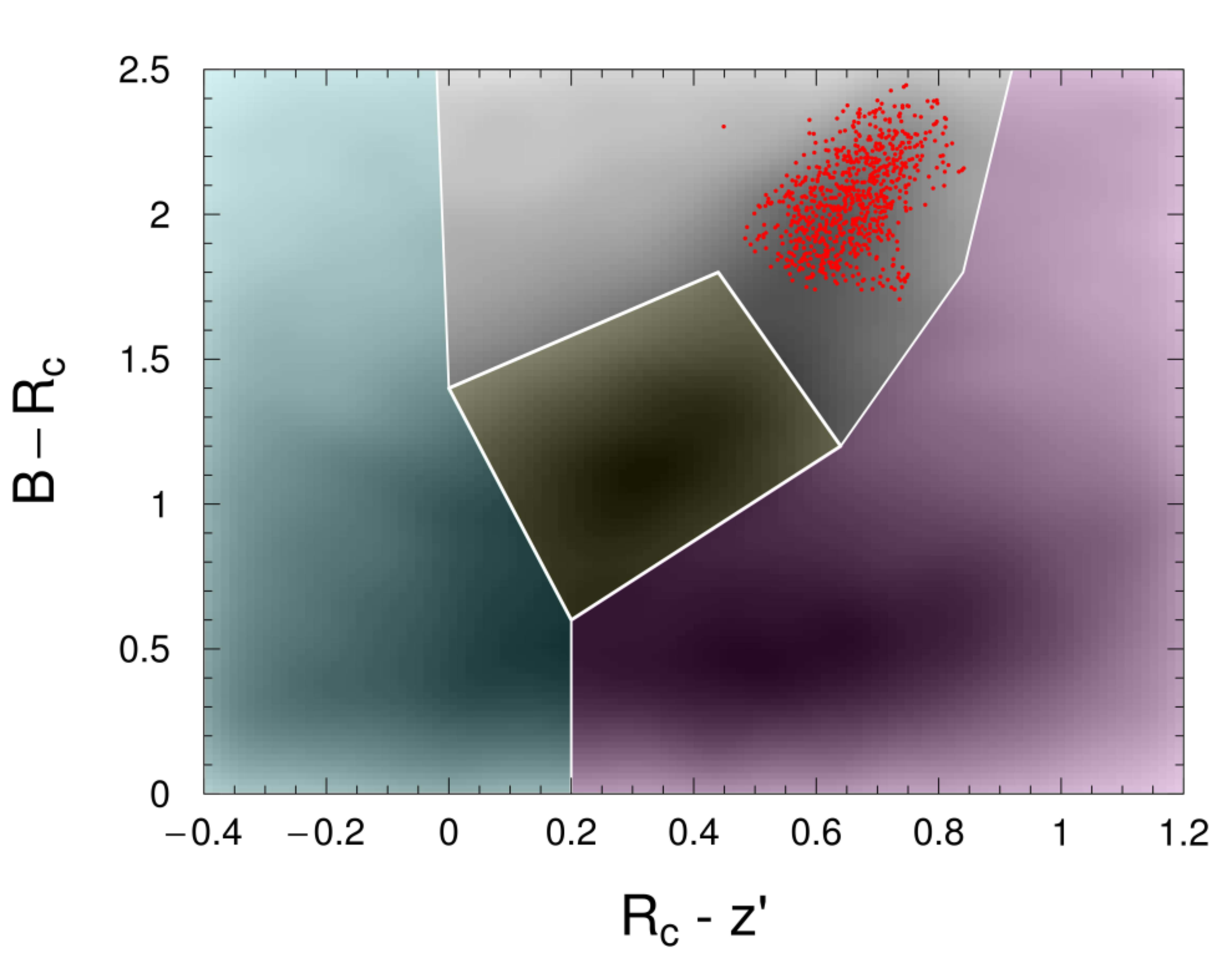} 
\caption[]{Identified galaxy populations in the colour-colour diagram. Using the criteria of \citet{med10} as guidelines we divided this space to select preferentially foreground (\textit{yellow}, centre), blue background (\textit{blue}, left), red background (\textit{magenta}, right) and, using statistical subtraction, red sequence cluster galaxies ({\it red} points). }
\label{fig:back}
\end{center}
\end{figure}

For the distortion sample we selected galaxies fainter than $R_C>22$ and within the ``background locus'' \citep[e.g.][]{capak07}
on a $(R_C-z') \times (B-R_C)$ colour-colour diagram (see Fig. \ref{fig:back}). The criteria used aimed first to minimize contamination by foreground and cluster galaxies and second to retain the maximum possible number of objects. After the shape measurement process (Sec. \ref{subsec:shear_measurement}), some of those galaxies are discarded based on the quality of the data.

\subsubsection{Shape measurement}
\label{subsec:shear_measurement}

In order to measure unbiased shapes of galaxies we need to map the atmospheric and instrumentation induced point spread function (PSF) over all the field of view and correct for it. This is done by using bright unsaturated stars that, by virtue of their negligible apparent sizes, appear PSF shaped on images. 

\begin{figure}
\begin{center}
\includegraphics[angle=-90, width=\columnwidth]{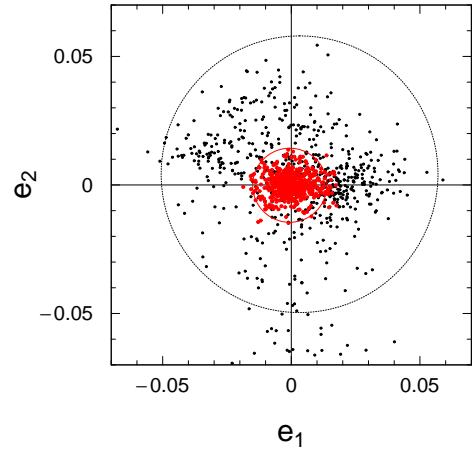} 
\caption[PSF measurement]{{\it Black} points: initial $e_1$,$e_2$ distribution with $\langle e_1\rangle=0.002$, $\sigma_{e_1}=0.023$ and $\langle e_2\rangle=0.003$, $\sigma_{e_2}=0.021$. If the stars were observed as real point sources we would expect  $\langle e_1^{\star} \rangle=\langle e_2^{\star} \rangle=0$ so the deviation is due to the PSF effect. {\it Red} points: residual distribution after the iterative process to quantify the PSF along the field with $\langle {\rm res}\rangle_{e_1}=0$, $\sigma_{{\rm res}_{e_1}}=0.006$ and $\langle {\rm res}\rangle_{e_2}=0$, $\sigma_{{\rm res}_{e_2}}=0.004$. The {\it circles} encloses $95\%$ of the points.}
\label{fig:elip}
\end{center}
\end{figure}

We performed the shape measurements of PSF deconvolution with the Bayesian code {\sc im2shape} \citep{im2shape}\footnote{\texttt{http://www.sarahbridle.net/im2shape/}} that models objects as a sum of Gaussians with elliptical basis. Stars are modelled as single Gaussian profiles thus no deconvolution is performed and the PSF dependent parameters in the model are kept: the ellipticity components $e_1$,$e_2$ and the FWHM. Using the thin plate spline regression \citep[{\sc Tps},][]{fields} in the {\sc R} environment \citep{R} we spatially interpolated our discrete set of PSF parameters to create a continuous function all over the field of view\footnote{We have adjusted $df=200$ parameters, leading to 346 degrees-of-freedom at the end. This procedure have resulted in smooth PSF parameter surfaces.}. We iterate this process three times, removing each time 10\% of the objects with larger absolute residuals. In Fig. \ref{fig:elip} we show the final measured ellipticities and corresponding residuals after the spatial interpolation.

Background galaxy shapes are then modelled as a sum of two Gaussians with identical basis and local PSF deconvolved. From this sample we remove objects with large ellipticity errors ($\sigma_{e}>2$) and with evidence of contamination by nearby objects. The final distortion sample, after measurements on the $z'$ band, had $12720$ objects or $\sim$24 gal. arcmin$^{-2}$. 
The use of the other two bands resulted in a smaller sample, mostly because of the worse image quality, therefore all the following analysis is based on the $z'$-band results only.

In order to translate the observational parameters into physical quantities we have to estimate and the average critical lensing surface density $\Sigma_{cr}$ for this particular ``distortion'' sub-sample. As equation \ref{eq:sig_crit} states, it requires some knowledge of the distribution of the source's redshift, given that we do not have (photometric) redshift estimations for each of our sources (the lens redshift is well known).
We circumvented this difficulty by using the COSMOS photometric redshift catalogue \citep{cosmos}, where we selected objects using the same criteria described before to select sources. Given the absence of $R_C$ in COSMOS, we calibrated between this magnitude at Subaru and COSMOS $g'$ and $r'$. Applying these procedures we found $\Sigma_{cr}=3.03(3)\times 10^{9}$ M$_\odot$ kpc$^{-2}$.

\subsubsection{Likelihood Estimation}
\label{subsec:shear_L_estimation}

Abell~1758 is a multimodal system, therefore we have to consider that each source galaxy has its image shape affected simultaneously by all the most important sub-structures. Moreover there is no circular symmetry; so, instead of dealing with the usual ``tangential'' component of the shear only, we have decomposed it into the Cartesian associated components 1 and 2. We do that by projecting the (tangential) effective shear in the Cartesian system by the lensing convolution kernel:
\begin{equation}
D_1 = \frac{y^2 - x^2}{x^2 + y^2}\, ,
\quad
D_2 = \frac{2xy}{x^2 + y^2} \, . 
\end{equation}
where $x$ and $y$ are the Cartesian coordinates in a space which has the respective lens centre at the origin.

Considering only the three main sub-structures of A1758, the effective shear components in each position of the lens plane can then be obtained by:
\begin{equation}
g_i = g_i^{\rm NE} +  g_i^{\rm NW} +  g_i^{\rm S},
\label{eq:g_sum_clumps}
\end{equation}
with $i=1,2$.

We can then estimate the misfit statistic $\chi^2_d$ (where the sub-index $d$ is after distortion) as:
\begin{equation}
  \chi^2_d=\sum_{j=1}^{N_{\rm sources}} \sum_{i=1}^{2}  \frac{(g_i-e_{i,j})^ 2}{\sigma_I^2+\sigma_{e_{i,j}}^2}, 
  \label{eq:chi2_d}
\end{equation}
where $\sigma_{e_{j_i}}$ is the measurement error given by {\sc im2shape} and $\sigma_I$ is the uncertainty associated with the intrinsic ellipticity distribution of the sources that, using our data, we estimated as $\sim 0.35$.

Assuming a Gaussian distribution of the data around the model, as usual, this statistics can be associated with the log-likelihood as:
\begin{equation}
\ln \mathcal{L}_d \propto - \frac{\chi^2_d}{2}. 
\end{equation}

\subsection{Magnification bias}
\label{subsec:kappa_measurement}

\subsubsection{Source selection and counts}
\label{subsec:mag_source_select}

For the ``magnification'' sample we used the red background galaxies (see Fig. \ref{fig:back}), following \cite{umetsu_broad11}. The blue background sample has been discarded as its number count has a logarithmic slope very close to $\alpha_{\rm blue} \sim 0.4$, which causes no observable magnification bias effects on the counts, see Eq.~(\ref{eq:bias_mag}). Using regions of our image away from A1758 structures ({\it blue} solid line in Fig.~\ref{fig:A1758} ) we measured $\alpha=0.10\pm 0.03 < 0.4$ for the red sample near the completeness limit, so we expect the lensing effect to cause a depletion of the number counts. We estimated the critical lensing density for this sample the same way we did for the depletion sample and found $\Sigma_{cr}=3.25(3) \times 10^{9}$ M$_\odot$/kpc$^2$.

In order to map the magnification bias effect we used the ``count-in-cells'' technique. We thus divided our image space in $68 \times 54 = 3672$ square cells with 15 arcsec on a side. We masked out regions occupied by large objects as saturated stars, foreground and cluster galaxies (the BCGs in particular). The area of each cell was computed after discounting the masked regions. Counts in regions away from the main structures (Fig. \ref{fig:A1758}) gave us a baseline number count of $N_0=35.7 \pm 13.4$ gal.~arcmin$^{-2}$.

\subsubsection{Likelihood Estimation}
\label{subsec:mag_L_estimation}

We can estimate the strength of the lensing signal by comparing the measured counts with the baseline number. Taking into account the appropriate uncertainties we can thus define a $\chi^2$ as:
\begin{equation}
  \chi^2_m=\sum_{i=1}^{N_{\rm cells}} \frac{(N_i-N_0~\mu^{2.5\alpha-1})^2}{\sigma_{N_0}^2}  \frac{W^2_i}{\sum_{j=1}^N{W^2_j}} \, ,
  \label{eq:chi2_m}
\end{equation}
where $N_i$ are the cumulative counts in each cell $i$, corrected by the effective (unmasked) cell area and $W = \sqrt{1-A_{\rm mask}/A_{\rm total}}$ is a weight that penalizes cells with small effective areas. The magnification $\mu$ is a model parameter here and depends on the shear and the convergence. The magnification from each sub-structure in our model is added linearly, as the effective shear (Eq. \ref{eq:g_sum_clumps}). 

Finally, using the same assumptions as before, the log-likelihood is 
\begin{equation}
\ln \mathcal{L}_m \propto - \frac{\chi^2_m}{2}. 
\end{equation}

We also assume in this analysis that the priors of $N_0$ and $\alpha$ are Normal distributions: $\Pi_{N_0} =  \mathcal{N} (N_0,\sigma_{N_0}^2/n)$, where $n$ here is the number of cells, and $\Pi_{\alpha} = \mathcal{N}(\alpha,\sigma_{\alpha}^2)$.

\subsection{Mass distribution modelling}
\label{sec:mass_model}

We have described the lensing observables in the A1758 field as the sum of three NFW \citep[][]{nfw96,nfw97} profile lenses \citep[see][for the gravitational lensing equations]{nfw_eq}, each associated with the three main sub-structures: NW, NE \& S.
The NFW profiles can be defined by two parameters: $M_{200}$ and the concentration parameter $c$, plus two others which locate its centre position ($x_c$, $y_c$) in a Cartesian plane. Therefore, the whole system would be defined by 12 parameters. We took some steps, however, to constrain this model.  

For all three lenses we adopted the prescription of \cite{duffy08}
\begin{equation}
c=5.71\left(\frac{M_{200}}{2\times10^{12}h^{-1}M_{\odot}}\right)^{-0.084}(1+z)^{-0.47},
\label{eq:duffy_rel}
\end{equation}
to relate the concentration parameter to the mass of the sub-structure. Moreover, for A1758S, we fixed the center of the mass distribution using its BCG position (see Fig. \ref{fig:A1758}). By doing that we reduced the number of free parameters to a more manageable number of
7, that focus on the problem at hand: $M_{200}^{\rm NE}$, $x_c^{\rm NE}$, $y_c^{\rm NE}$, $M_{200}^{\rm NW}$, $x_c^{\rm NW}$, $y_c^{\rm NW}$ and $M_{200}^{\rm S}$.

Adopting a Bayesian approach, we included in these set of parameters two quantities that describe the counts of the unlensed populations: $N_0$ and $\alpha$. Those ``nuisance parameters'' will be fitted along the sub-structure related parameters but we established Normal priors for both, based on our measurements (Sec. \ref{subsec:kappa_measurement}). For all the masses, we used a flat prior in the range $0<M\leq 6\times 10^ {15}$ M$_{\odot}$ that was useful only to increase the convergence speed by cutting off unrealistic high masses. We set no (informative) priors at all for the centres of the two Northern structures.

Calling the vector of NFW parameters as $\theta$ we can write the posterior of our problem as:
\begin{multline}
\noindent \mathcal{P}(\theta,N_0,\alpha|{\rm data}) \propto  \\
\mathcal{L}_d({\rm data}|\theta) \times \mathcal{L}_m({\rm data}|\theta,N_0,\alpha)~\Pi(N_0)~\Pi(\alpha)\, .
\end{multline}
\label{eq:posterior}

The posterior defined here carries information coming from both image distortion and number count depletion data. One can easily define posteriors for individual techniques by just removing the elements not related to them off this expression.

\subsection{Results}
\label{sec:lensing_results}

\subsubsection{Masses}
\label{sec:mass_results}

We have used the MCMC (Markov chain Monte Carlo) algorithm with a simple Metropolis sampler to map this posterior through the {\sc MCMCmetrop1R} \citep[][]{MCMCpack} that runs within the {\sc R} environment \citep[][]{R}. We run four chains of $10^5$ points with different seed each after $10^4$ ``burn-in'' iterations to ensure that the chains starts into the stationary state. We have calculated the potential scale reduction factor $R$ as proposed by Gelman \& Rubin \citep{coda} to control the convergence of the MCMC output and found that for all ($7+2$ in the most general case) parameters $R<1.1$ within $68\%$ c.l., ensuring non divergence of the chains. We run models using data from both techniques (d+m),  distortion only (d) and magnification only (m).

\begin{table}
\caption[]{Marginalized masses in $10^{14}$ M$_{\odot}$ unities. Averages and 68\% range of the MCMC samples obtained using distortion only ``d'', magnification only ``m'' and both techniques ``d+m''.}
\label{tab:masses}
\begin{center}
\begin{tabular}{|l|ccc|}
\hline \hline
 & M$_{200}^{\rm NW}$        & M$_{200}^{\rm NE}$        & M$_{200}^{\rm S}$ \\
\hline

d         &$10.32_{-2.33}^{+2.44}$& $5.14_{-2.12}^{+1.37}$& $5.01_{-1.47}^{+1.29}$\\[5pt] 
m         & $6.00_{-3.77}^{+1.88}$& $7.81_{-3.55}^{+2.86}$& $5.62_{-2.54}^{+1.76}$\\[5pt] 
d+m       & $7.90^{+1.89}_{-1.55}$& $5.49^{+1.67}_{-1.33}$& $4.96^{+1.08}_{-1.19}$\\
\hline \hline
\end{tabular}
\end{center}
\end{table}

\begin{figure}
\begin{center}
\includegraphics[angle=0, width=\columnwidth]{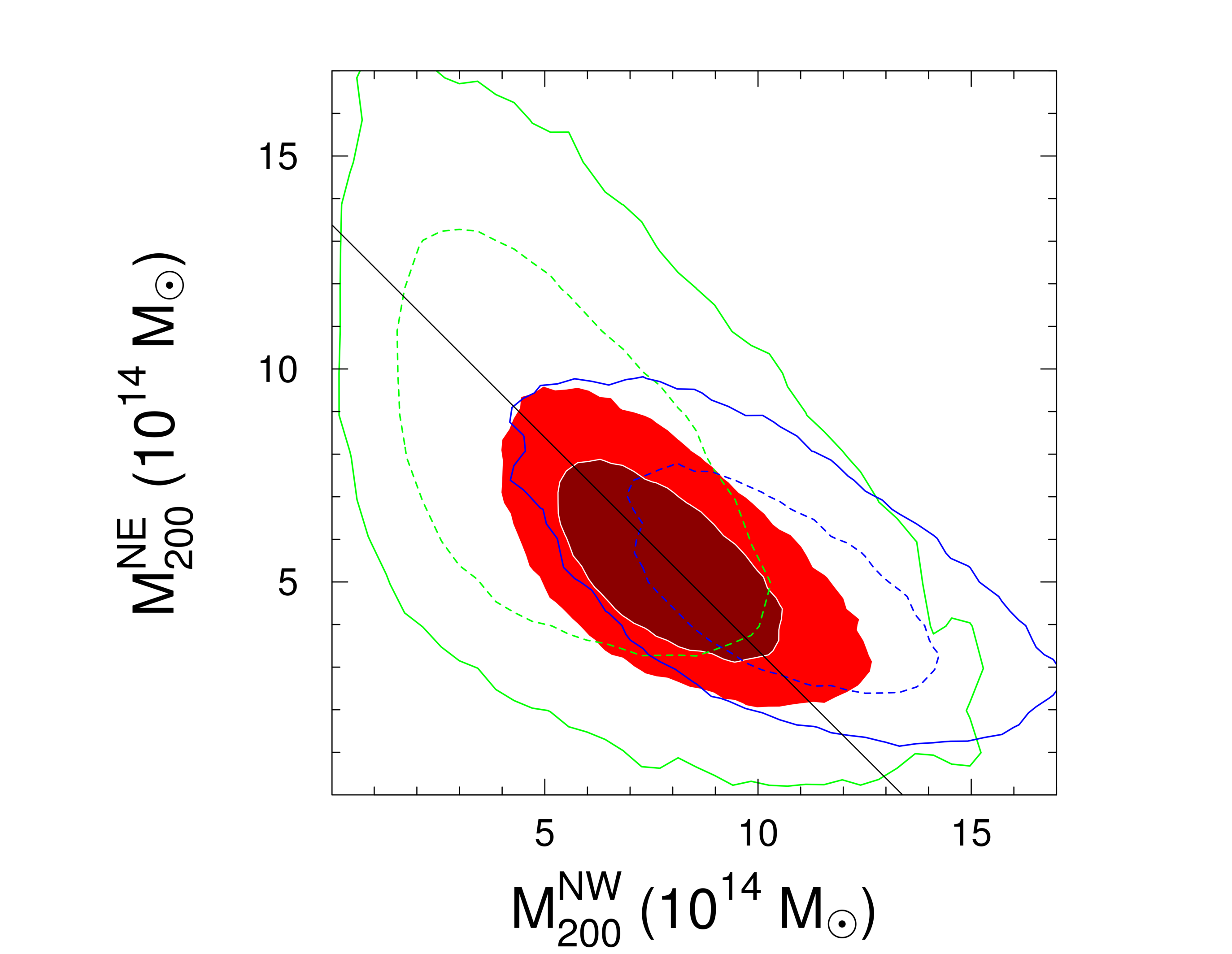} 
\caption[]{Marginalized $2\sigma$ posterior of the Northern masses NW \& NE for distortion only (d, \textit{blue}), magnification  only (m, \textit{green}) and combined data (d+m, \textit{red}). Despite a large mass degeneracy  we can observe a good match among them and specially between d and d+m datasets,with the last one being the most precise result. The  \textit{black} line corresponds to the sum of the masses ($M^{\rm NW}_{200}+M^{\rm NE}_{200}$) and it is  nearly parallel to the posteriors main axes.} 
\label{fig:mass_posterior}
\end{center}
\end{figure}

In Table \ref{tab:masses} we show the average values of the MCMC samples for the three clump masses and respective 68\% intervals. 
The posteriors for Northern masses, marginalized over all other parameters, are shown in Fig. \ref{fig:mass_posterior}. 
From both table and figure, it can be seen that the independent distortion and magnification measurements are consistent with each other, having significant overlap of their 68\% c.l. regions. This is fortunate because justifies the joint use of both techniques. 
Distortion alone, as expected, clearly has more power to constrain mass parameters, showing 68\% confidence intervals which are $\sim 67\%$ smaller than magnification alone. The combination of both decreases the confidence intervals to $\sim 80\%$ with respect to distortion alone.

\begin{figure*}
\begin{center}
\includegraphics[angle=90, width=1.0\textwidth]{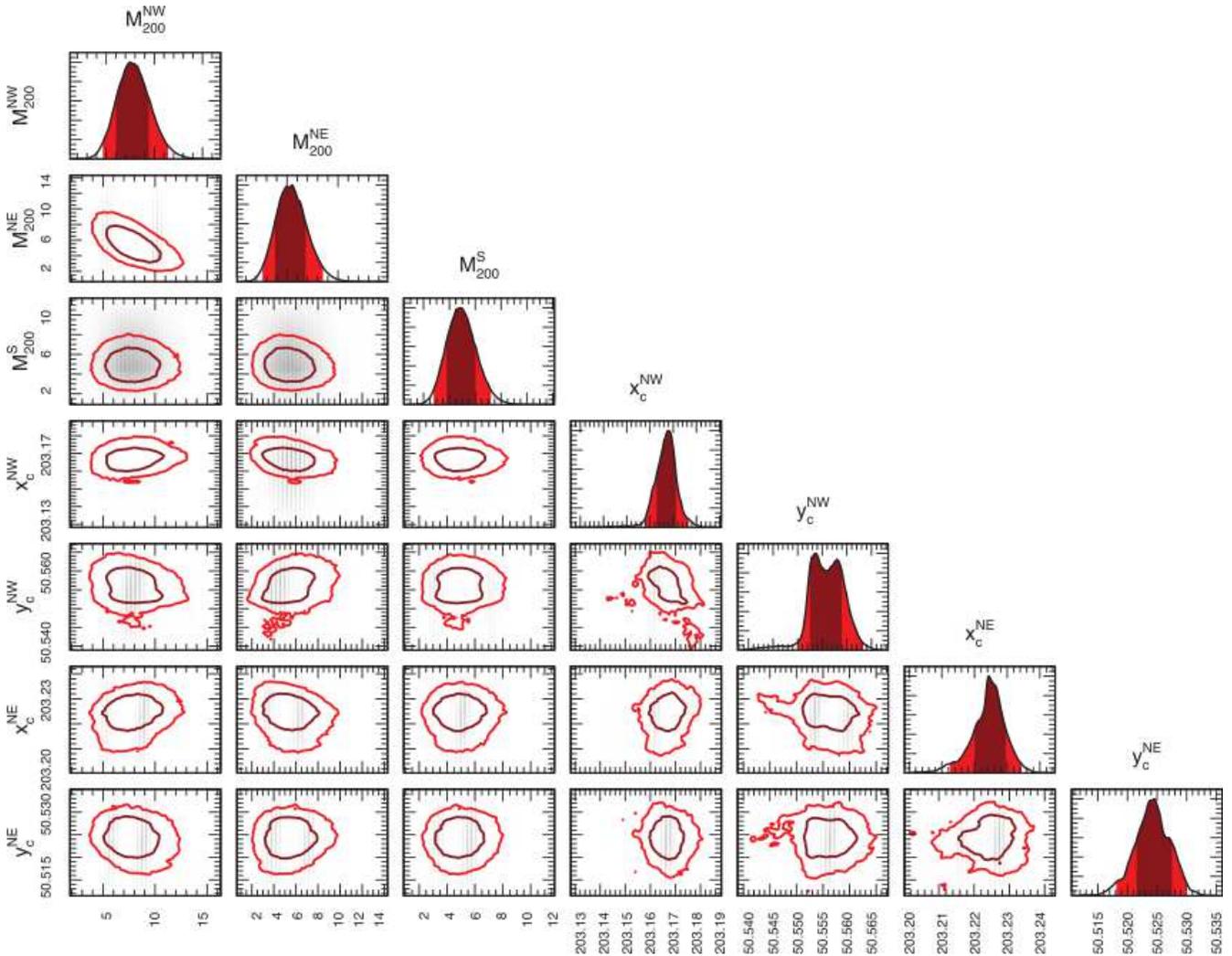} 
\caption[]{Posteriors of all 7-NFW parameters for the  combined distortion and magnification data (d+m). In the \textit{diagonal} we can see the marginalized individual posterior of each parameter whereas the correlations are presented in the lower triangle showing the confidence curves for $1 \sigma$ and $2 \sigma$. A high absolute inclination of the shape denotes a higher correlation between the two considered parameters, as we can see, for example, between the Northern masses. All masses are in $10^{14}$ M$_{\odot}$ units, whereas the positions are shown in the world coordinate system (degrees). } 
\label{fig:triangle}
\end{center}
\end{figure*}

\begin{figure}
\begin{center}
\includegraphics[angle=0, width=1.0\columnwidth]{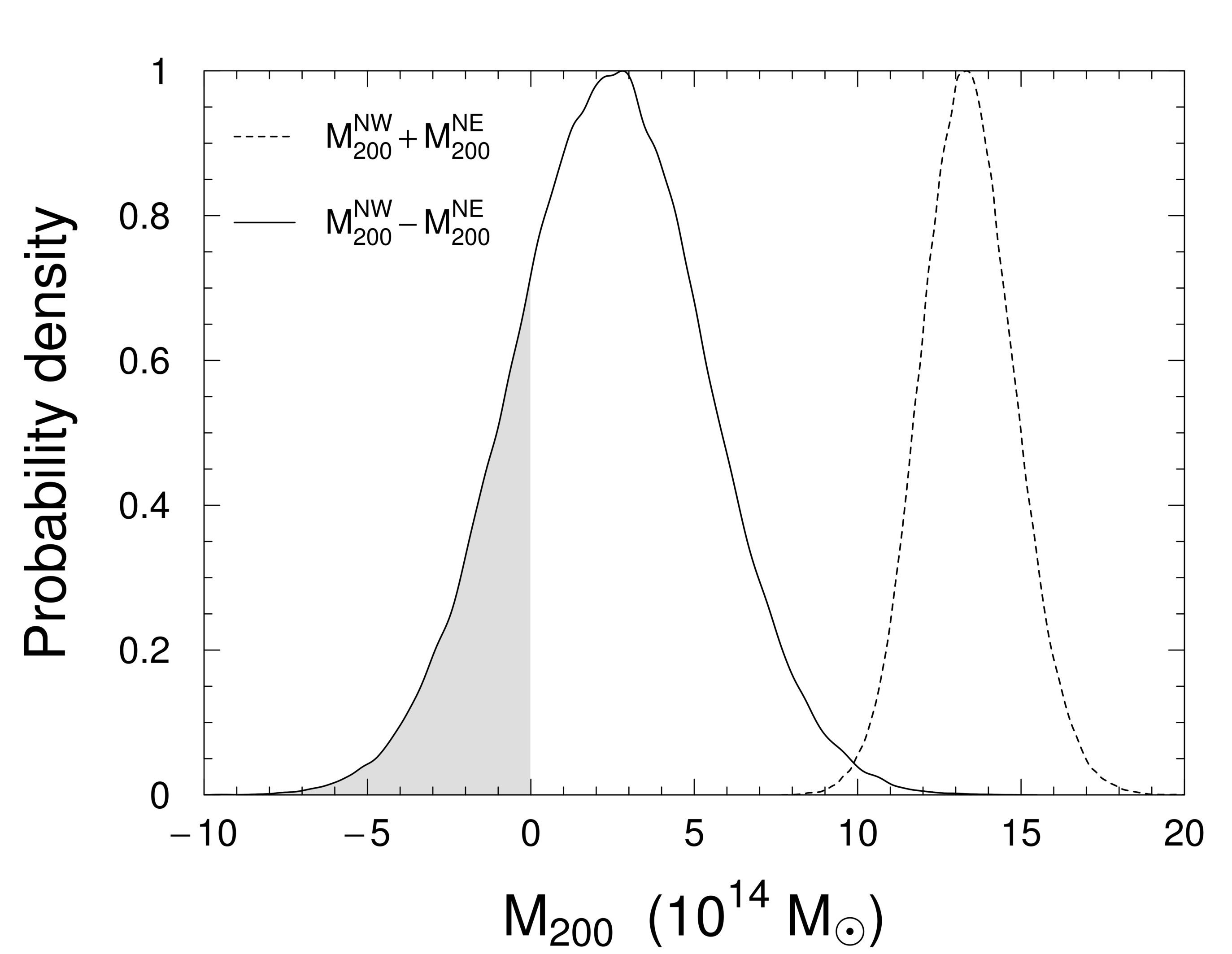} 
\caption[]{Probability distribution for the subtraction of the masses (NW--NE, {\it continuous} line) and  their sum ({\it dotted} line). Both distributions have been normalized to peak at one. The grey region, 21\% of the total, is related to the probability that $M^{\rm NE}_{200} > M^{\rm NW}_{200}$.}
\label{fig:sub_mass}
\end{center}
\end{figure}

It can also be seen in Fig.~\ref{fig:mass_posterior} that there is an important degeneracy between NW \& NE masses. Actually, among all parameters, those two are the most correlated, as we can see in Fig. \ref{fig:triangle} where we show the posteriors of all 7 parameters (d+m). In fact, it can be noticed that the main degeneracy axis is nearly parallel to constant  masses (continuous line in the plot). To test that, we looked into the density distributions of the sum and difference of the two Northern masses (Fig. \ref{fig:sub_mass}).

As it can be seen, our data strongly constrains the sum, whose posterior can be well summarized as $M_{200}^{\rm NW}+M_{200}^{\rm NE} = 13.39^{+1.37}_{-1.45} \times 10 ^{14}$ M$_\odot$, but constrains more poorly the difference among them, $M_{200}^{\rm NW}-M_{200}^{\rm NE}= 2.41\pm2.93 \times 10^{14}$ M$_\odot$. In other words, our data and analysis do a good job measuring the mass of A1758N, but perform less well in splitting this mass among the two sub-structures \citep[][had similar problems]{ragozzine}. It is of interest, however, that  there is a probability of $\sim79\%$ that A1758NW is the more massive of the two (Fig. \ref{fig:sub_mass}). This is relevant because it may help explain why this structure shows no signs of gas detachment.

We also checked whether or not there is any trade-offs between the masses of A1758N (NW+NE) and A1758S. The Spearman correlation coefficient between them is only $\rho=-0.128 \pm 0.003$, thus there is little to none degeneracy between the mass chains of the two A1758 main components. 

\subsubsection{Central positions}
\label{subsub:positions}

As mentioned before, one of the main goals of this work is to determine the central position of the clumps NW \& NE, so we can attach a more precise estimate of significance to the offsets between peaks of X-ray emission and the actual centre of the mass halo.

\begin{figure}
\begin{center}
\includegraphics[angle=-90, width=\columnwidth]{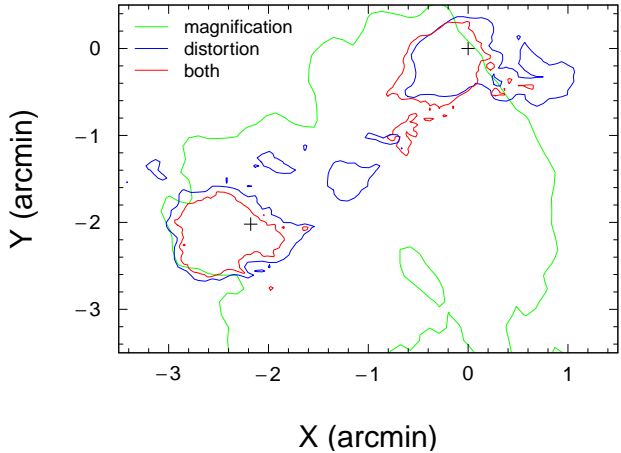} 
\caption[]{Confidence curves (95\% c.l.) of the NW \& NE centre positions for each data set used in this work: magnification (m, {\it green}), distortion (d, {\it blue}) and combined (d+m, {\it red}). The BCG NW was considered the origin of our coordinate system.  {\it Crosses} identify the BCGs.} 
\label{fig:positions}
\end{center}
\end{figure}

We first consider the constraints from the week-lensing analysis. The location of the substructures central positions, described in the previous section, are presented in Fig.~\ref{fig:positions}, with contours of the 95\% confidence areas for the posteriors of the central position parameters of A1758NW/NE, marginalized over all other parameters. As we can clearly see, magnification alone (green contours) produces very poor constraints, therefore the combined results are very close to those of distortion alone.

In order to determine the X-ray peaks positions, we have downloaded the publicly available data from the Chandra archives\footnote{\texttt{http://cda.harvard.edu}}. There are five pointings: 7710 (PI: Gordon P. Garmire), 2213, 13997, 15538, and 15540 (PI: Laurence David) in A1758 field. We have reprocessed and combined them into a single broad band [0.3--7.0 keV], exposure-map corrected image using \textsc{ciao}~4.6.8, following the standard procedure described in the CXC ``Science Threads''\footnote{\texttt{http://cxc.cfa.harvard.edu/ciao/threads}}. The isophote contours from the resulting image are shown in Fig.~\ref{fig:A1758}.

The peaks of the X-ray emission were determined by using the barycentre technique. It consists on a search for local maxima, around which we define a box of  $0.6 \times 0.6$ arcmin ($\sim$ the size of central brightest regions). The peak is then the intensity weighted average position.

\begin{figure*}
\begin{center}
\includegraphics[angle=0, width=0.8\textwidth]{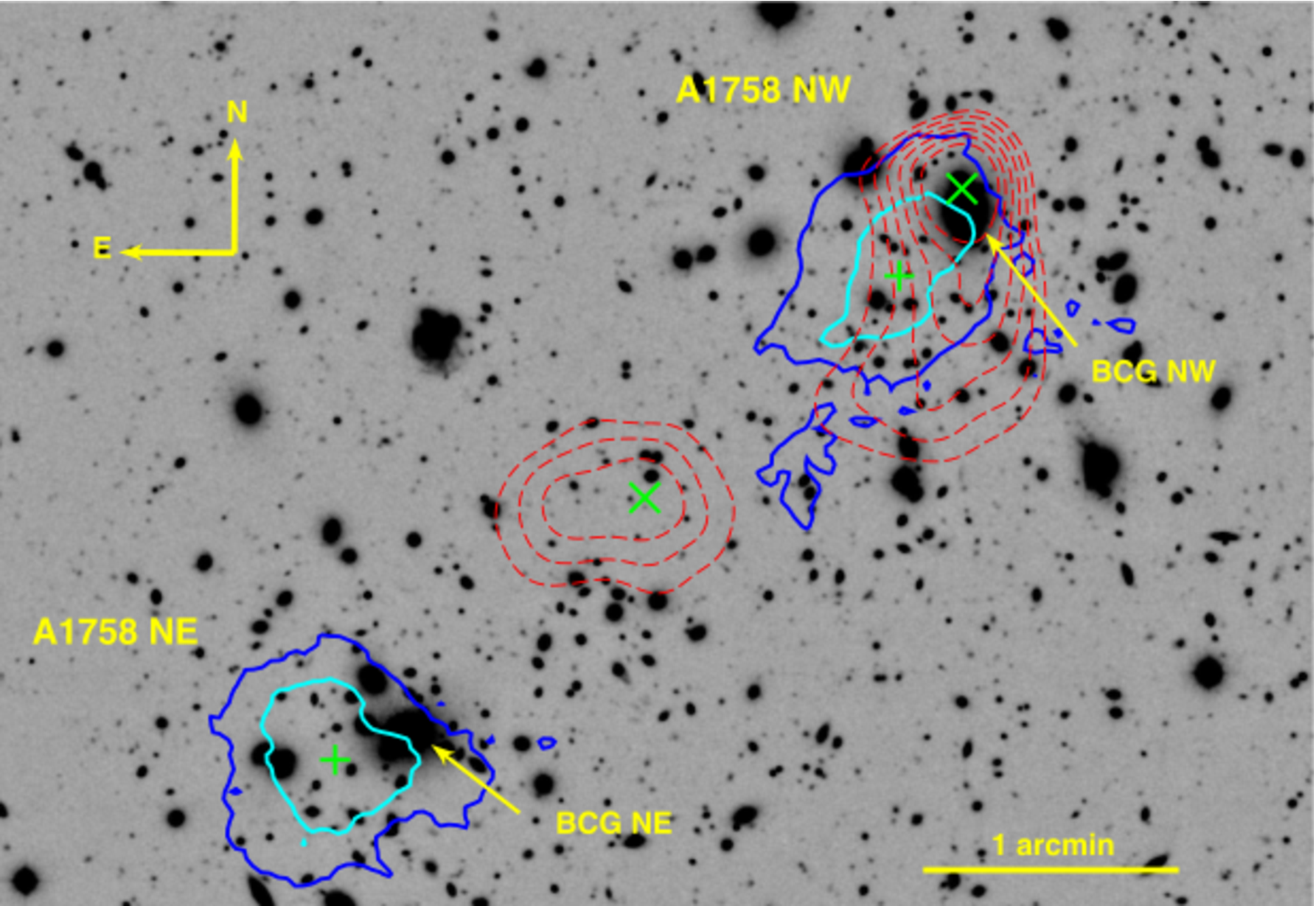}  
\caption[]{Combined optical $z'$ band, innermost ICM distribution traced by X-ray emission observed by Chandra ({\it red}), $1\sigma$ ({\it cyan}) and  $2\sigma$ ({\it blue}) confidence contours of  NW \& NE mass centre positions. The ``+" marks the mean position for the mass centres  and the ``X"  shows the X-ray peak locations, each of them related with the nearest subcluster (NW, right  \& NE, left). While the BCG NE position is compatible with the respective mass centre within $2\sigma$, the X-ray peak shows a considerable detachment ($96_{-15}^{+14}$ arcsec) for the last one, as we can expect for post-merger systems.  On the other hand, in NW the positions of the three components (BCG, mass peak and  X-ray peak) are compatible within $2\sigma$.} 
\label{fig:result}
\end{center}
\end{figure*}

In Fig. \ref{fig:result} we show a closer view of A1758N, where we can better compare the BCG positions, X-ray distribution and mass centre location.
As in previous studies \citep{sevenmergers,ragozzine} we found no significant offsets between the BCGs and the mass centres. Those distances are $24_{-14}^{+10}$ and $30_{-18}^{+12}$ arcsec, for NW and NE, respectively. Thus, in both cases,  the BCG position is consistent with the halo mass centre within 95\% c.l.

In Fig.~\ref{fig:result} one can also see that the position of the NE clump centre is  $96_{-15}^{+14}$ arcsec distant from the corresponding X-ray peak. Qualitatively similar results were  reported by \citet{sevenmergers} and \citet{ragozzine}. A separation as large as this  cannot be explained by the presence of fore/background structures, shear shape noise, nor NFW model artifacts \citep{dietrich12}. For A1758NW, the position of the X-ray peak emission is compatible with that of the BCG and the mass clump centre, considering the uncertainties. A visual inspection of the Fig.~2 in \cite{harvey15} has shown that about 25\% of their merging clusters presented the  configuration where the detachment of the ICM from the other cluster components, as observed in the Bullet cluster, is seen only in one of the components.

\section{Dynamical analysis}
\label{sec:dynamical_analysis}

In this Section we will describe the analysis of the dynamics of A1758 member galaxies based on their radial velocities. For that we used literature data plus new spectroscopic data we describe below. 

\subsection{Spectroscopic data} 
\label{subsec:spec_data}

Spectroscopic observations were done in 2010A with the Gemini Multi-Object Spectrograph (GMOS) mounted on the Gemini North telescope. We used five slit masks (four targeting A1758N and one A1758S) with a total combined of 186 slitlets. Galaxies with $R_C < 22$ from the red sequence in a $(B-z) \times z$ space were priority targets, but we also targeted some bluer galaxies to  maximize the number of objects per mask. For each mask the total integration time was 66 minutes. We used the R400 grating and 1.0 arcsec wide slits, which produces spectra with  $\Delta \lambda \approx$  8 \AA~ in 6500 \AA.  Multi-slit standard reduction and calibration in wavelength was done with {\sc gemini.gmos IRAF} package. 

Radial velocities for galaxies with absorption lines were obtained by the cross-correlation technique \citep{td79} using the task {\sc xcsao} from {\sc RVSAO} package \citep{rvsao}. We have correlated the data with several galaxy spectra templates of similar resolution but large S/N. For each template the code outputs a redshift. To select among them we used the \cite{td79} cross-correlation coefficient $r$ as our primary criterion, complemented by a visual inspection of the spectra, superimposed with the positions of the main spectral lines. This last step is important to reject some wrong solutions found  occasionally in low S/N spectra. As a final criterion, we discarded all data with $r \geq 3$. In 14 galaxy spectra we identified emission lines. In those cases the redshifts were obtained by their analysis (task {\sc emsao}). In the end of this processes we had 165 reliable redshifts: 130 for A1758N  and 35 for A1758S.

The current sample has 62 galaxies in common with \cite{boschin}. The residual between estimated radial velocities of 60 out of those 62 objects is $\langle v_{\rm our}-v_{\rm B12}\rangle=-26$ km s$^{-1}$, with a dispersion of $\sigma_{\langle \rm res \rangle}=195$ km s$^{-1}$. 
We have excluded the galaxies \#108 and \#119 (respectively \#89 and \#119 in \cite{boschin}) from this analysis for having residuals larger than $5\times10^3$ km s$^{-1}$. After careful visual inspection of both spectra we decided to keep our own redshift measurements as the fiducial ones for these objects. Two galaxies from the \cite{boschin} sample (\#38 and \#115) had their images blended with other objects in our data, which compromised the photometric data, and thus were discarded.

\begin{figure*}
\begin{center}
\includegraphics[angle=-90, width=0.4\textwidth]{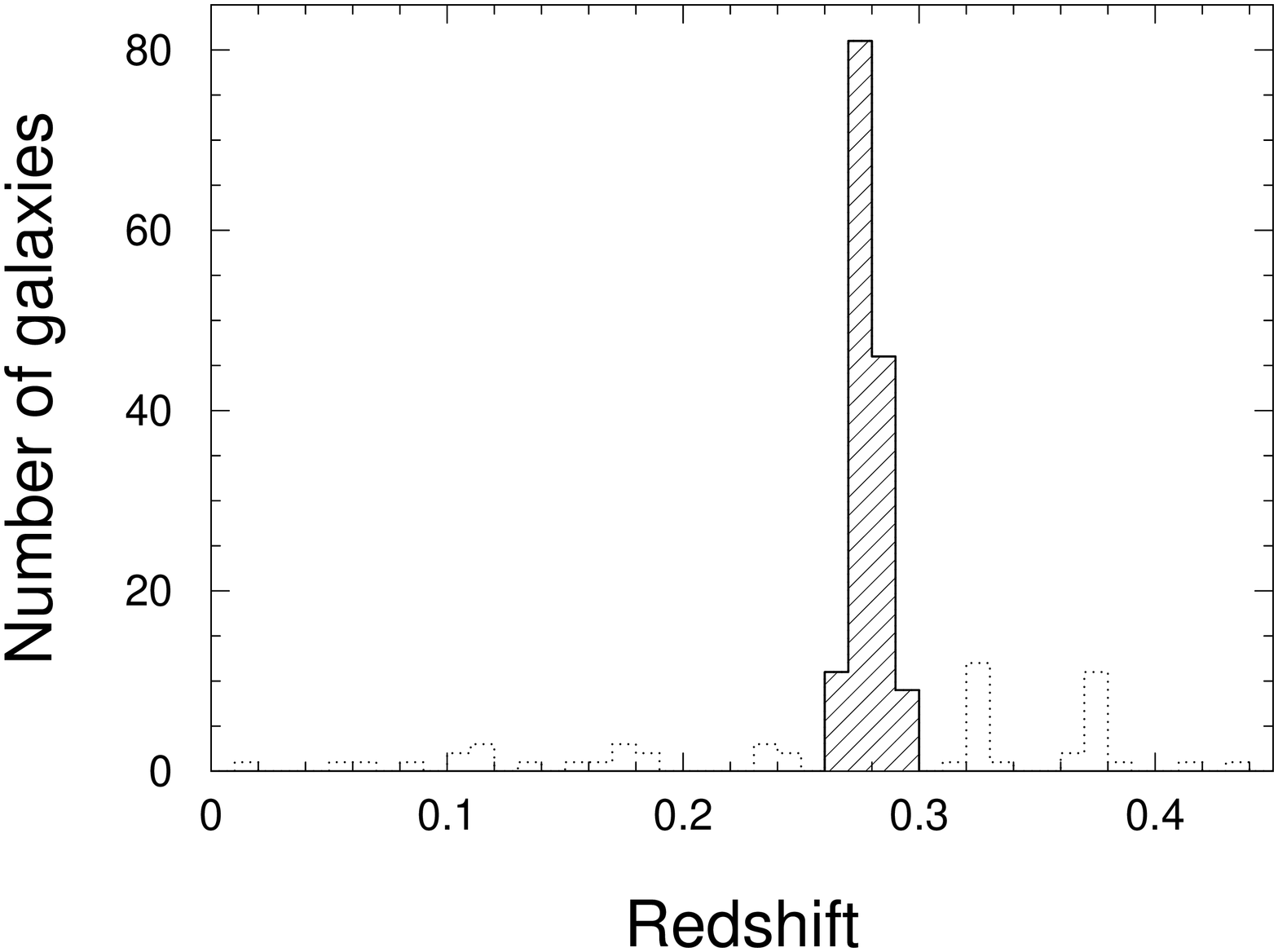} \quad 
\includegraphics[angle=-90, width=0.4\textwidth]{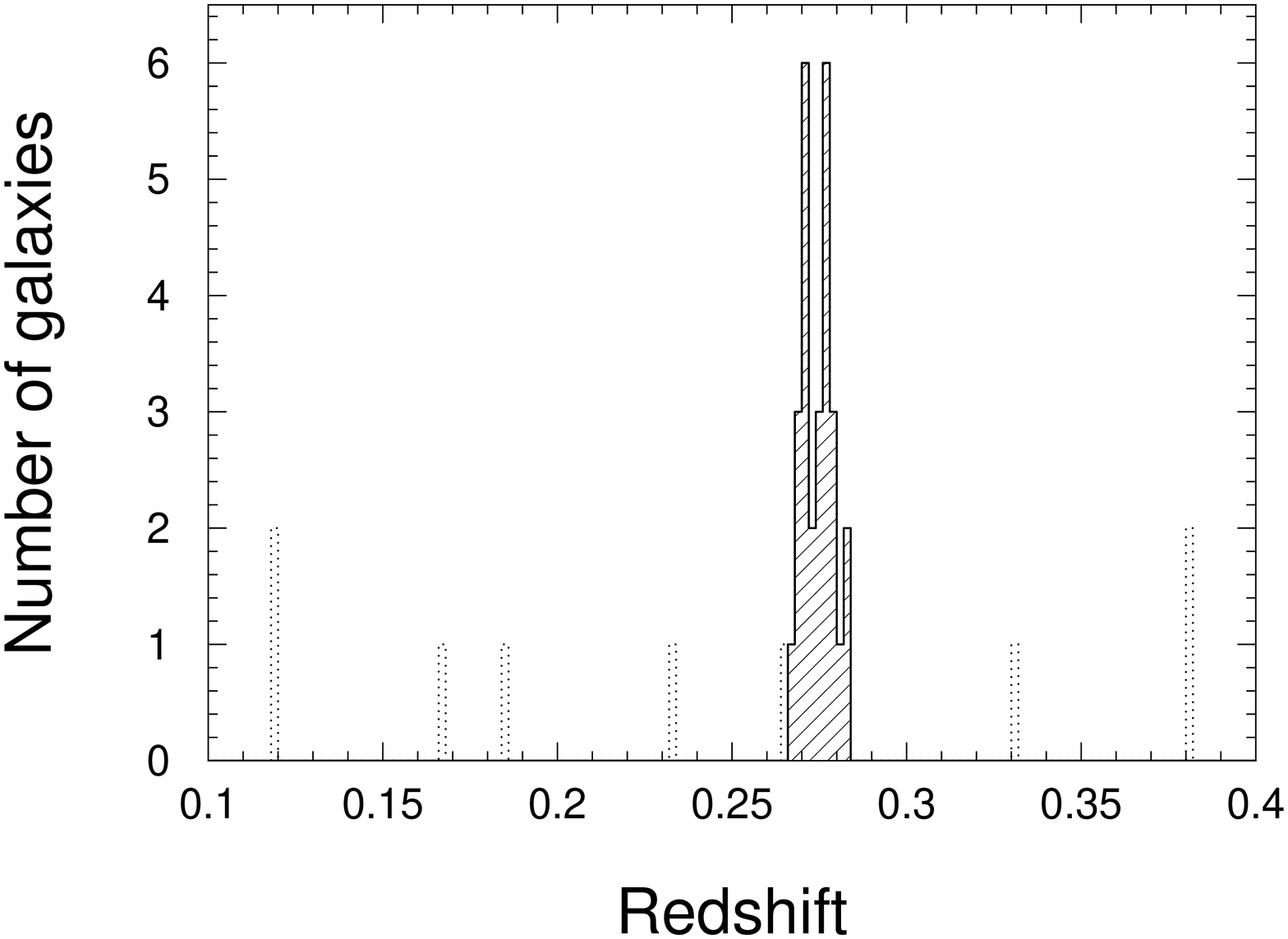} 		   
\caption[]{\textit{Left: }Redshift distribution in A1758 Northern field highlighting the 147 galaxies selected as probable cluster members  (\textit{dashed}) after $3\sigma$-clipping. Four galaxies with $z\geq0.45$ were excluded for clarity. \textit{Right: } The same for the A1758 Southern field where we selected 27 galaxies as cluster members.} 
\label{fig:vdistall}
\end{center}
\end{figure*}

Our final catalogue of A1758N field (see Appendix \ref{ap:A1758Ncatalogue}) has 203 galaxies: 70 are new measurements presented by us here, 73 by \cite{boschin},  plus 60 from both catalogues with consistent measurements. For the A1758S field (see Appendix \ref{ap:A1758Scatalogue}) we present 35 new redshifts.  The velocity distribution in both A1758 fields can be seen in Fig. \ref{fig:vdistall}.

In order to remove outliers from the velocity distribution of each sample (North and South), we  used the 3$\sigma$-clipping method \citep{3sigmaclip} iteratively, until each sample mean did not vary by more than 1\%. This procedure resulted in the identification of 147 members centred in $\bar{z}=0.279$ with dispersion  $\sigma_v/(1+z)=1442$ for A1758N  and  $\bar{z}=0.274$~km~s$^{-1}$ and $\sigma_v/(1+z)=1221$ km s$^{-1}$ for the 27 galaxies belonging to A1758S. The final redshift distributions can be seen in Fig. \ref{fig:vdistall}.

\subsection{A1758N}
\label{subsec:dynamical_A1758N}

\subsubsection{Identification of substructures}
\label{subsub:A1758N_3d}

More subtle substructures (e.g. infalling galaxy groups) are in general not identified by the  3$\sigma$-clipping procedure as they present velocities consistent with those of the general cluster population. However, the member galaxies of those substructures are clustered in both projected space and velocity spaces, and this can be used to unveil them.

There is a large amount of substructure tests but, according to \cite{pinkney}, the Dressler-Shectman (DS or $\Delta$) test \citep[][Eqs. \ref{eq:ds} and \ref{eq:sumds}]{ds} is the most sensitive when 3D data (radial velocities plus projected positions) are available and thus we used it. It works by quantifying the deviation of the local ($N_{\rm nb}=\sqrt{N}$ closest neighbours, where $N$ is the total number of galaxies in the cluster) systemic velocity and dispersion to the those of the overall structure.
%
\begin{equation}
 \delta_i=\left \{ \left ( \frac{N_{\rm nb}+1}{\sigma^2} \right ) [(\bar{v}_{l}-\bar{v})^2+(\sigma_l - \sigma)^2] \right \}^{1/2}
 \label{eq:ds}
\end{equation}

\begin{equation}
\Delta=\sum_{i=1}^{N} \delta_i
 \label{eq:sumds}
\end{equation}

When analysing a relaxed cluster one expects $\Delta \sim N$ \citep{ds} whereas a substructured one will have $\Delta \ge N $. In an complementary approach, based on mock catalogues, \cite{hou12}  argued that a p-value $<0.01$ is the better index for substructure identification, minimizing both false negative and false positive DS-test results. This p-value was calculated as follows: we have generated $n$ samples and in each one we have kept the positions fixed, shuffled the velocities and applied the DS test generating $\Delta_{s}$ to compute   
\begin{equation}
\textrm{p-value} = \frac{\sum(\Delta_{s} > \Delta)}{n} \,,
 \label{eq:pvalue}
\end{equation}
where we have counted in how many realizations $\Delta_{s}>\Delta$. 

\begin{figure}
\begin{center}
\includegraphics[angle=-90, width=\columnwidth]{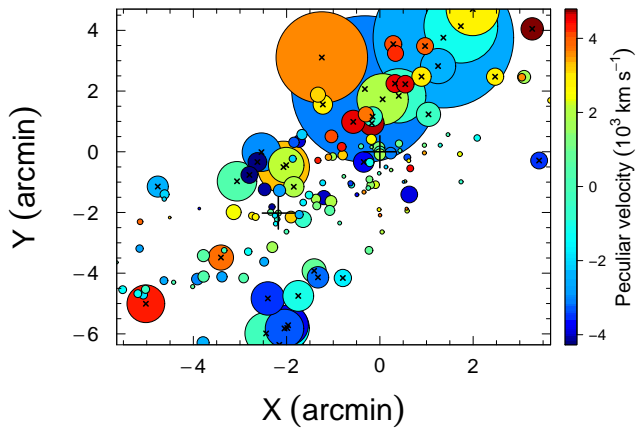}            
\caption[]{Result of DS test. {\it{Circles}} indicate galaxy positions and their sizes are proportional to $e^{\delta_i}$ in the sense that the larger this value, the more relevant is the substructure. Colors indicate the galaxy peculiar velocity $v_i-\bar{v}$ and  ``X" marks galaxies belonging to some substructure according to our $\delta_i>2$ criterion (see text for more details). The BCGs are marked as ``+".} 
\label{fig:ds}
\end{center}
\end{figure}

Our measurements with $N=147$ galaxies ($N_{\rm nb}=12$) found $\Delta=225.6$ within $99\%$ significance and a p-value$=4.6\times 10^ {-4}$ ($10^5$ resamplings). Both figures point to the presence of substructures, which can be seen in the ``bubble plot'' shown in Fig. \ref{fig:ds}.

%

To identify and remove these substructures we searched  for a threshold $\delta_c$ (Eq. \ref{eq:ds}) below which a galaxy $i$ is considered part of some substructure. For each $\delta_i \in \{\delta_1,...,\delta_N\}$ we remove all galaxies having $\delta$ greater than  $\delta_i$ and we calculated a $\Delta$ statistic for the remaining sub-sample. After this process, it was clear that a  $\delta_c \approx 2$ works as a good threshold and thus we kept in our sample all galaxies with $\delta$ lower than this value. The remaining ``main structure'' with 105 galaxies (hereafter A1758N cluster members) is the sample that will be used hereafter.

\subsection {The NE-NW velocity difference}

One crucial information that is required to fully characterize the dynamical state of the A1758NE-NW system is the radial velocity difference between those two sub-clusters. Below we present two attempts to divide the A1758N members between NE or NW. One using the radial velocity data alone (1D) and another using the projected positions (2D).

\subsubsection{1-D analysis}
\label{subsub:A1758N_1d}

We tested the radial velocity distribution using several normality tests. All of them fail do rule out a uni-modal (DIP) or a Normal (A-D, Shapiro, D'agostino, Anscombe) distribution within 99 \% c.l., as none of the p-values are smaller than $0.15$ (See table \ref{tab:stests}).

 \begin{table}
  \caption[]{Summary of the normality tests applied on the A1758N redshift distribution.}
 \begin{center}
 \begin{tabular}{c c}
 \hline \hline 
 Test name & p-value \\
  \hline 
 DIP $^{a,d}$			       &$ 0.99$ \\
 Anderson Darling $^b$  & $0.15$\\
  D'agostino  $^c$         & $0.99^e$\\
 Anscombe  $^c$          & $0.89^f$\\		 
  \hline \hline
 \multicolumn{2}{l}{$^a$ \cite{dipR}.} \\
 \multicolumn{2}{l}{$^b$ \cite{nortest}.} \\
 \multicolumn{2}{l}{$^c$ \cite{moments}.} \\
\multicolumn{2}{l}{$^d$ The DIP test is actually a uni-modality test.} \\
\multicolumn{2}{l}{$^e$ Measured skewness  was $-9\times10^{-4}$. }\\
\multicolumn{2}{l}{$^f$ Measured kurtosis was $2.92$.}\\ 
 \end{tabular}
  \label{tab:stests}
 \end{center}
 \end{table}

On a further attempt to detect an east-west bimodality in the velocity distribution of A1758N we tried two different methods that aim to find a given number of clusters in data-sets. Those are the  normal mixture modelling through expectation-maximization implemented in the R package {\sc Mclust}  \citep{mclust} and  partition around ``medoids''  \citep[PAM,][]{pam}. Those are methods of different approaches. {\sc Mclust} is model-based as it models the data distribution as a sum of Gaussians, whereas PAM is a partitioning method which creates subsets of data by minimizing the Euclidean distance, in the data space, between data points and central (``medoid'') positions. 

\begin{figure}
\begin{center}
\includegraphics[angle=-90, width=\columnwidth]{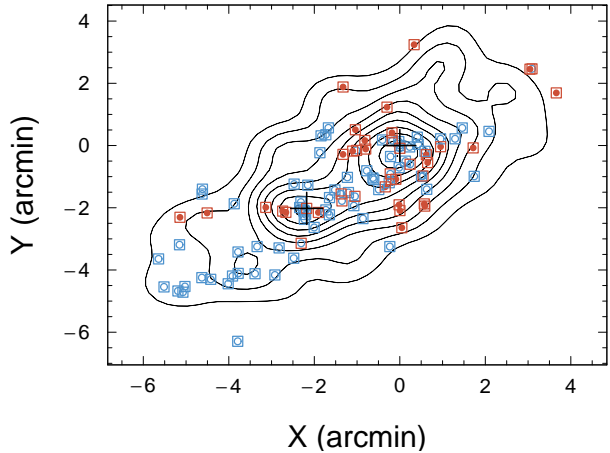}   
\caption[]{Spatial distribution of the sub-groups recovered by PAM ({\it circles}) \& {\sc Mclust} ({\it squares}) 1-D algorithms overlaid with the smoothed projected distribution of the photometric selected red sequence galaxies ({\it black} lines, see more details in Section \ref{subsub:A1758N_2d}).  The 1-D algorithms assign the majority of galaxies to similar groups but fail to assign the BCGs (``+'') to different ones.  Both sub-groups are spread along the field showing no signs of the bimodality as seen in the red sequence galaxies positions.  The first one ({\it blue} symbols) has $\bar{z}_1\approx 0.276$ whereas the second  one ({\it red} symbols) has $\bar{z}_2\approx 0.286$ for PAM and $\bar{z}_2\approx 0.281$ for {\sc Mclust} }.
\label{fig:1Dresults}
\end{center}
\end{figure}

In Fig. \ref{fig:1Dresults} we show the projected positions of the sub-groups created by PAM and {\sc Mclust}. In both cases, when dividing the sample in two sub-groups, independent of the inputs details, the east-west bimodality has not been reproduced.  In particular, both methods fail to separate the BCGs which we consider, {\it a priori}, to be the main representatives of each A1758N substructures. The PAM method, moreover, produced velocity distributions that resemble truncated Normal distributions that are unexpected, even for merging structures.

\subsubsection{2-D analysis}
\label{subsub:A1758N_2d}

\begin{figure*}
\begin{center}
\includegraphics[angle=-90, width=0.4\textwidth]{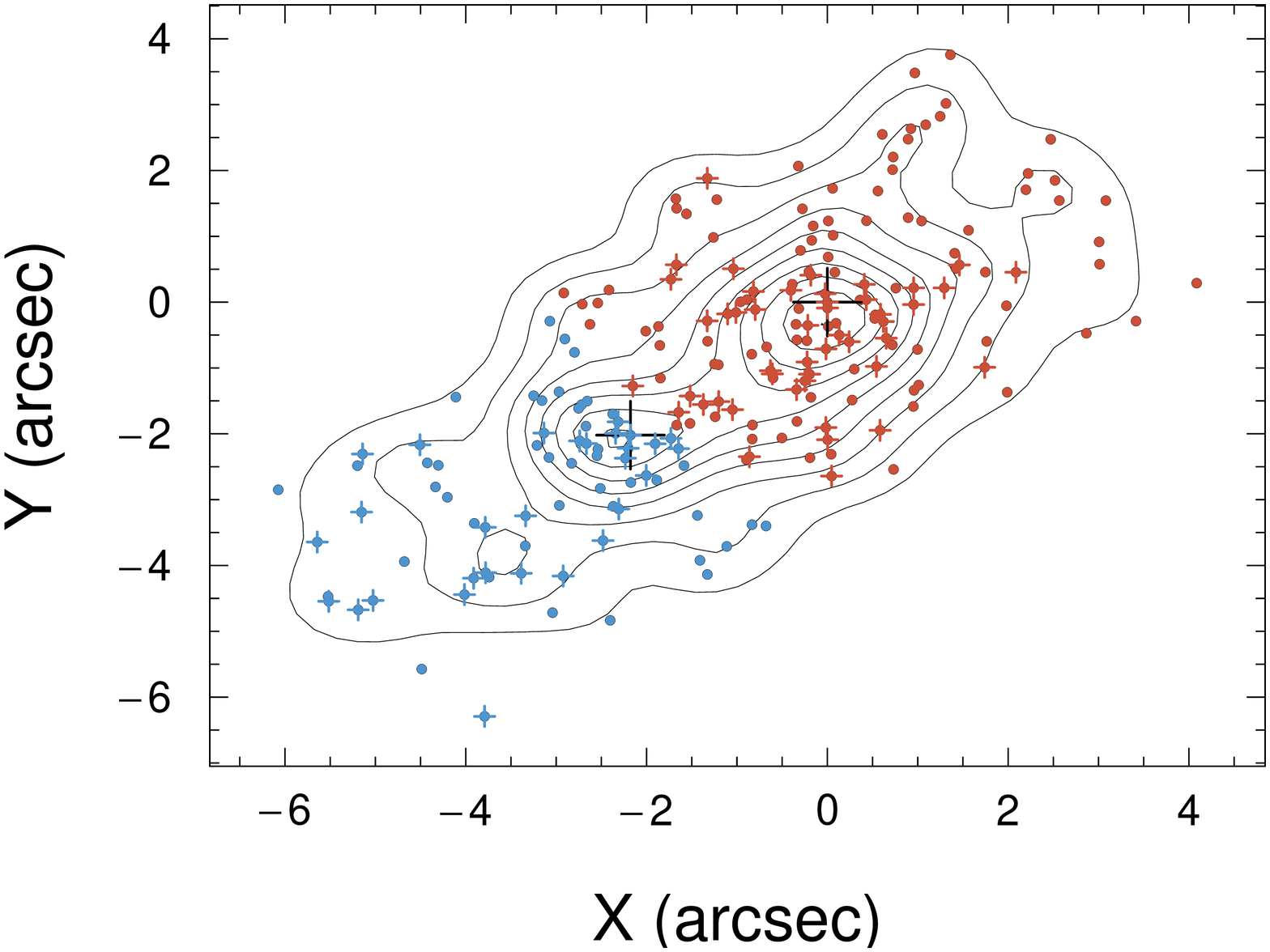}\quad 
\includegraphics[angle=-90, width=0.4\textwidth]{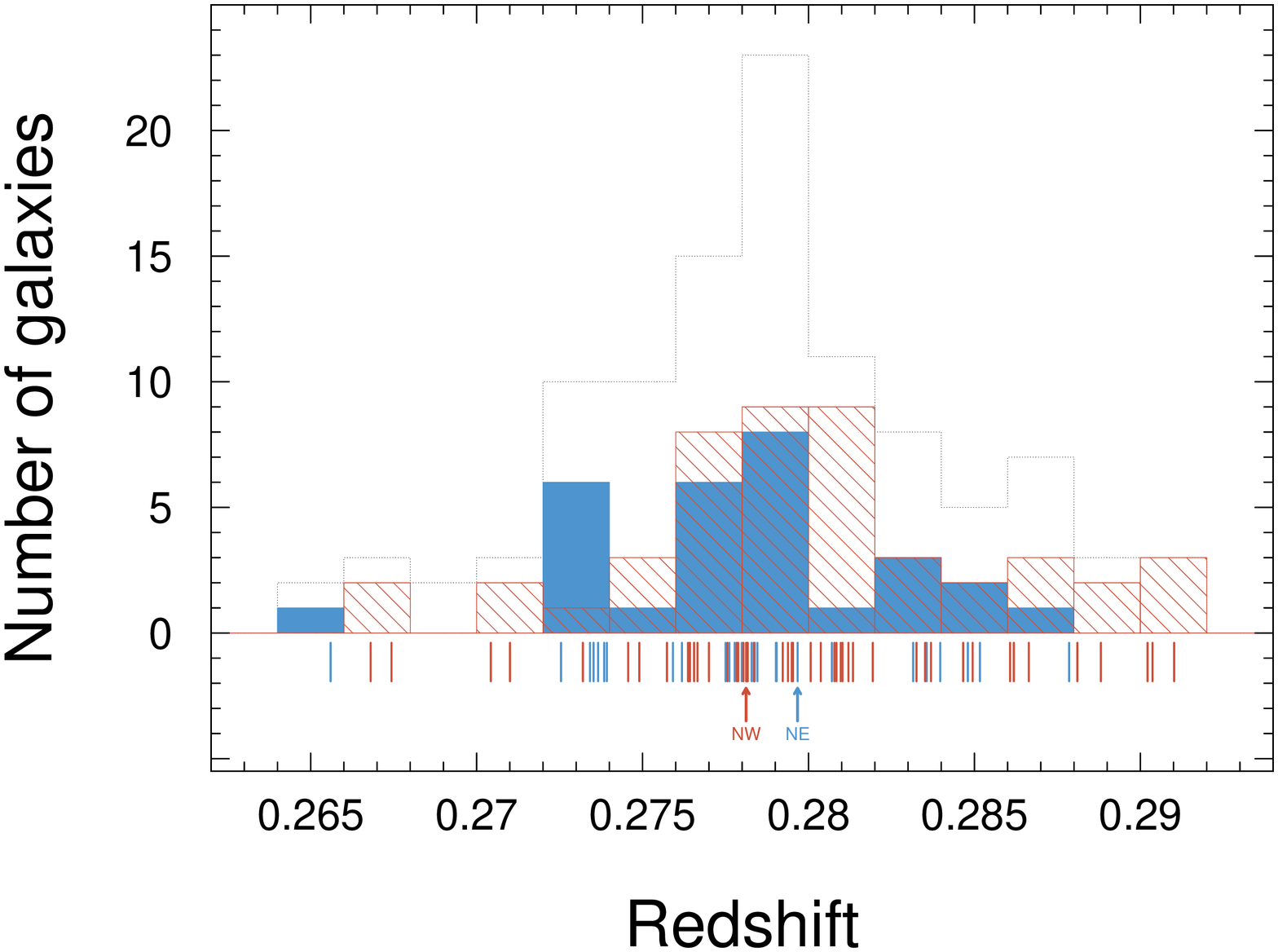}		     
 \caption[]{\textit{Left:}  Spatial distribution of the recovered groups A1758NW \& A1758NE (\textit{red} and \textit{blue} dots, respectively) overlaid with the smoothed density of the red sequence galaxies (\textit{black} lines). Crosses indicate the galaxies where the redshift are available and the \textit{black} ones indicate the BCGs positions. \textit{Right:} Redshift distribution of the recovered groups, characterized by  $\bar{z}_{\rm NE}= 0.2780(7)$, $\sigma_{\rm NE}/(1+z)= 1075$ km s$^{-1}$ and $\bar{z}_{\rm NW}= 0.2799(6)$, $\sigma_{\rm NW}/(1+z)= 1296$ km s$^{-1}$.  \textit{Upper arrows} show the BCGs redshifts.} 
\label{fig:2d1d}
\end{center}
\end{figure*}

Given that we cannot detect any east-west bimodality as observed in X-rays and mass in the velocity distribution of member galaxies we will turn to their projected distribution in the sky.  For that we will use all spectroscopically confirmed member galaxies plus all objects from the cluster red sequence up to $ R_{C}= 22$ in the area covered by our spectroscopic masks. Those were the primary source for spectroscopic targets and the results showed that most of them (69\%) were indeed members. We use the same methods previously described and the results are summarized in Fig.\ref{fig:2d1d}.

The results now show a clear east-west separation in the projected distribution, with groups that can be associated with the aforementioned A1758NE and A1758NW subclusters, where each BCG lies in a different group and with entangled velocity distributions. This is all in line with the expectations of a post-core passage collision, not far from the plane of the sky, as suggested by \cite{davidkempner} and \cite{Machado+2015}.
 
For the {\sc Mclust} method using the simplest ellipsoidal model EEE (equal volume, shape, and orientation) \citep[see][for more details]{mclust}, we obtained for the A1758NE and A1758NW subclusters, respectively $\bar{z}_{\rm NE}= 0.2780(7)$, $\sigma_{\rm NE}/(1+z)= 1075$ km s$^{-1}$ and $\bar{z}_{\rm NW}= 0.2799(6)$, $\sigma_{\rm NW}/(1+z)= 1296$ km s$^{-1}$. This gives a  difference in velocity of $\delta v/(1+z) = 423\pm275$ km s$^{-1}$. 
PAM results are compatible within $1 \sigma$ with those of {\sc Mclust} and its different covariance structures \citep[a detailed description of the {\sc Mclust}'s covariance models could be seen in][]{mclust}.
Moreover, the results are consistent even when using only core members (within 1.3 arcminute form each BCG).

\subsection{A1758S}
\label{subsec:dynamical_A1758S}


After the $\sigma$-clipping process already described, our A1758S sample ends up with 27 spectroscopically confirmed members, all new measurements (see right panel of Fig. \ref{fig:vdistall} and  Appendix \ref{ap:A1758Scatalogue} for the complete catalogue). The DS test founds no evidence for substructures producing $\Delta=38$ and a p-value equals to $0.03$ for $10^4$ resamplings. Applying the same tests we used for the  Northern structure, we also found that uni-modality and normality of the velocity distribution cannot be discarded within 99\% c.l.  This last result is in apparent contradiction with the claim of \citet{davidkempner} that A1758S is a merging system happening not far from the line-of-sight.

Given \citet{davidkempner} X-ray based results, we investigated the possibility that A1758S could be composed of two systems with similar projected distributions but different velocity distributions. For that we have applied the same statistical analysis tools used with the A1758N sample (Section \ref{subsub:A1758N_1d}) forcing them do find two sub-structures. The results are presented in Table \ref{tab:A1758Sresults}.

 \begin{table}
  \caption[]{Summary of the tests applied to A1758S data.}
 \begin{center}
 \begin{tabular}{l c c c c}
 \hline \hline 
  & $\bar{z}_1$ & $\frac{\sigma_1}{1+z}$ & $\bar{z}_2$ & $\frac{\sigma_2}{1+z}$ \\
    & 	 & (km $s^{-1}$) &  &  (km $s^{-1}$) \\
  \hline
 {\sc PAM} 			 & $0.2702(5)$	&  $527$  & $0.2783(5)$   &   $588$ \\
 {\sc Mclust 1-D}  & $0.2703(5)$ 	&  $565$  & $0.2782(6)$   &   $616$\\
 {\sc Mclust 3-D}  & $0.2713(8)$	&  $877$  & $0.2772(8)$   &   $877$ \\ 
  \hline \hline
 \end{tabular}
  \label{tab:A1758Sresults}
 \end{center}
 \end{table}
 
\begin{figure*}
\begin{center}
\includegraphics[angle=-90, width=0.4\textwidth]{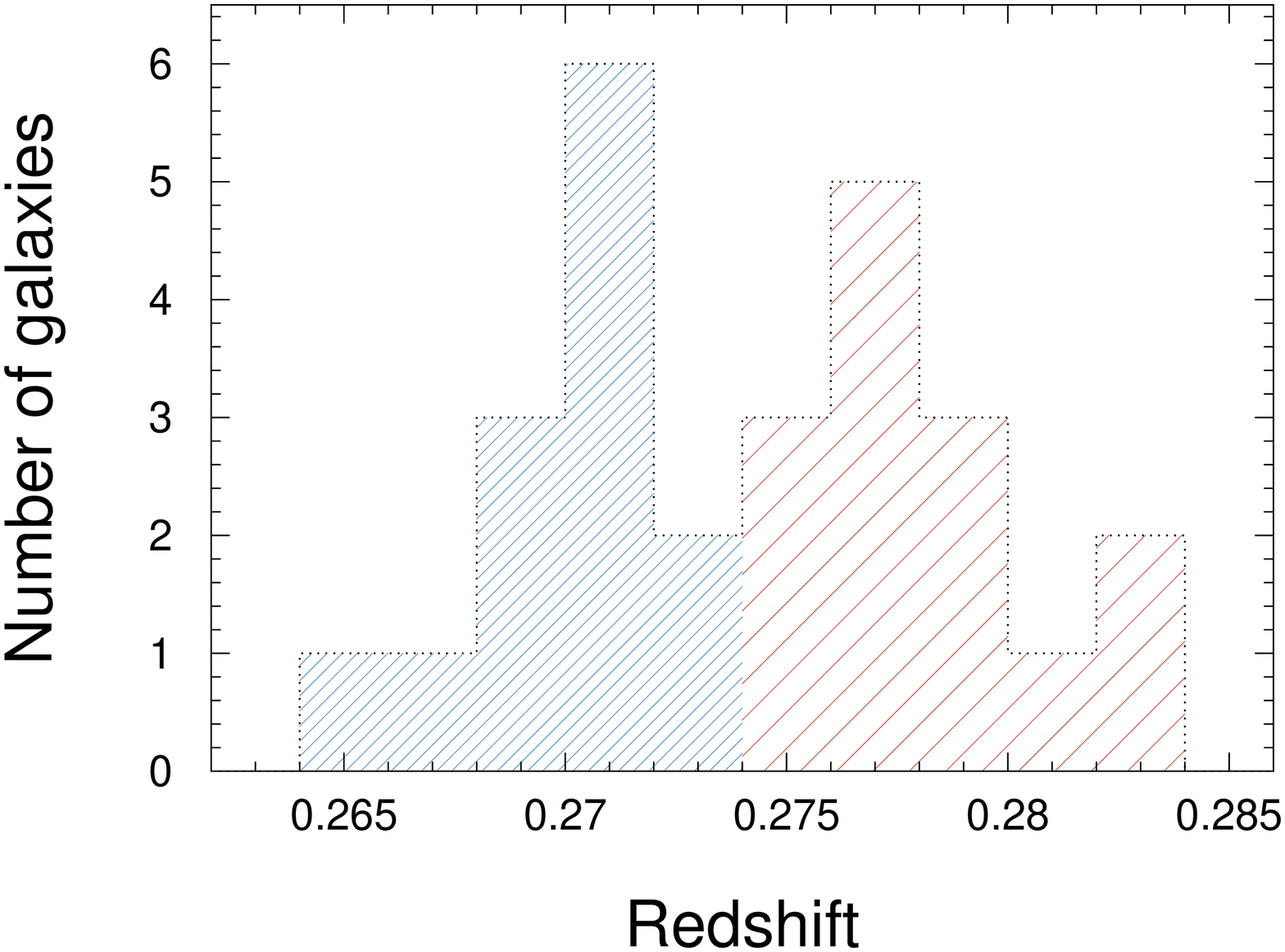}\quad 
\includegraphics[angle=-90, width=0.4\textwidth]{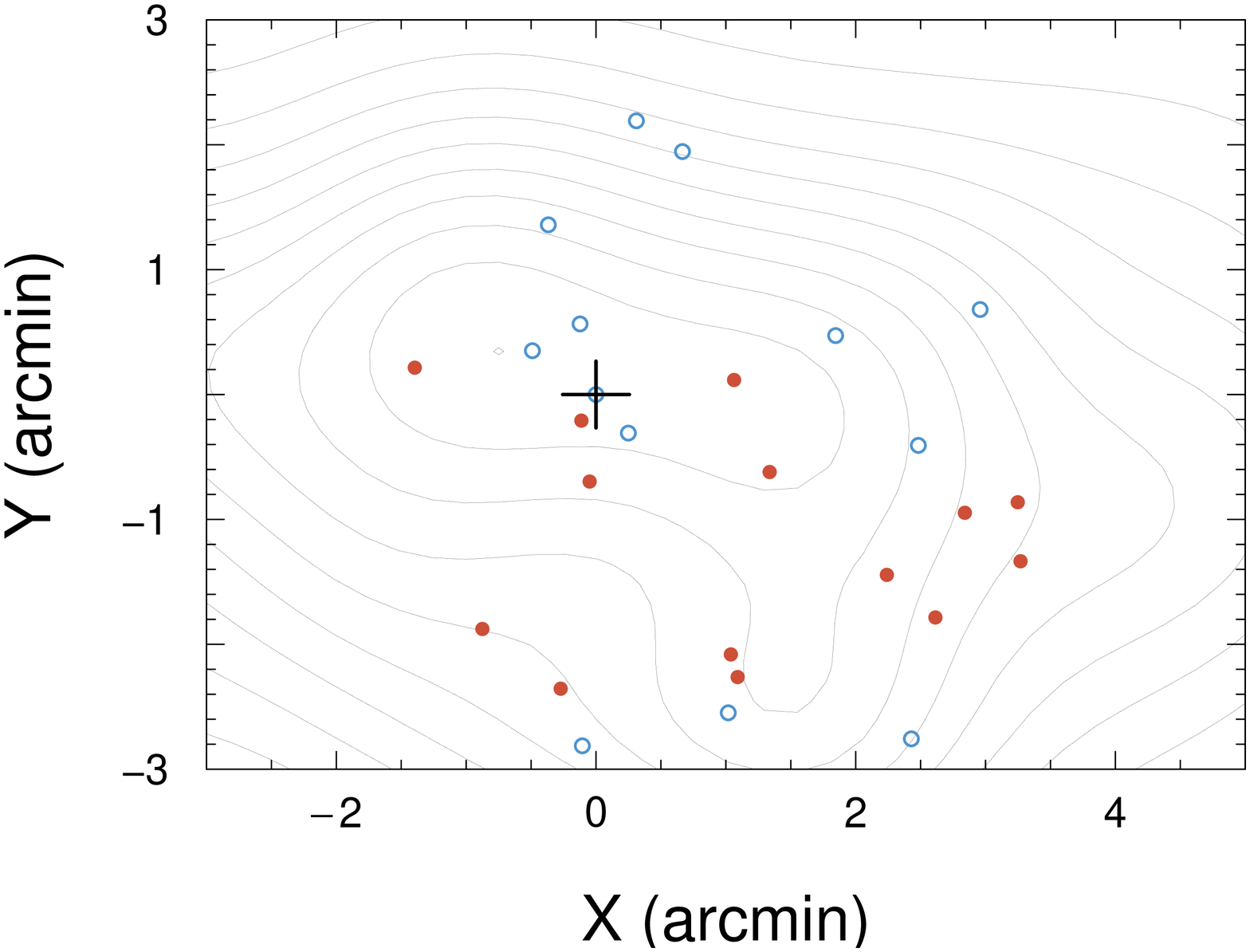}         
\caption[]{{\it Left: }Histogram of PAM \& {\sc Mclust 1-D} modelled sub-groups for the Southern region of A1758. According {\sc Mclust 1-D}   the {\it blue} sub-group with 13 members  has $\bar{z}=0.2703(5)$ and $\sigma /1+z=565$ km $s^{-1}$ whereas the 14 {\it red} members have  $\bar{z}=0.2782(6)$ and  $\sigma /1+z=616$ km $s^{-1}$. The velocity difference is $\delta v/(1+z)=1843\pm228$ km $s^{-1}$. {\it Right: }Spatial position of previously identified groups overlaid with red sequence galaxy position contours. Southern BCG is designated as ``+''.} 
\label{fig:pam_histS}
\end{center}
\end{figure*}

In general all tests agree within $1 \sigma$, but the 3-D test shows systematically higher values. We have chosen {\sc Mclust 1-D} as our fiducial result, since it presents modelled values (instead of PAM which shows data measures) with lower errors than its 3-D counterpart. Redshift distribution and spatial distribution are showed in Fig.  \ref{fig:pam_histS}. The results point into a large line-of-sight separation between the two sub-groups $\delta v/(1+z)=1843\pm228$ km $s^{-1}$. 

Considering a velocity of $700\pm100$ km s$^{-1}$ in the plane of the sky obtained through ICM density jumps by \cite{davidkempner}, plus considerations on the difference between the shock wave and the actual dark matter clumps \citep[][]{springel07}, we can estimate that the main merger axis has an angle of  $\theta=70\pm4$ degrees (see Eq. \ref{eq:angle} below) with respect to the plane of the sky.

\section{Updated hydrodynamical simulations of A1758N}
\label{sec:simulations}

Having these new mass estimates -- $(7.9 \pm 1.6) \times 10^{14} M_{\odot}$ and $(5.5 \pm 1.6) \times 10^{14} M_{\odot}$ for the A1758NW and A1758NE substructures respectively -- we updated the hydrodynamical $N$-body simulations  described in \citet[][see Tab.~\ref{tb:ic}]{Machado+2015}. The simulations are carried out using the smoothed particle hydrodynamics (SPH) $N$-body code {\sc Gadget-2} \citep{Springel2005}. Here we report on one simulation model that succeeded in reproducing the global morphology of A1758N, using the improved mass estimates of this paper.

The first pericentric passage takes place at $t=1.25$ Gyr. At that instant, the separation between the dark matter centroids is approximately 80 kpc. While this distance may seem small compared to cluster scales, even such modest pericentric separations are known to bring about sizable asymmetries in cluster collisions \citep[e.g.][]{Machado2015}. The observed morphology is approximately achieved at $t=1.52$ Gyr, i.e. 0.27 Gyr after the first pericentric passage.

Overall, these simulation results do not differ dramatically from the model presented in \cite{Machado+2015}. In that early model, we had explored a collision of two clusters having nearly equal masses. We had in that way established that it is, in principle, possible to obtain the very specific dissociative morphological features of A1758N from such a major merger. Here, using improved mass estimates derived from the gravitational weak lensing analysis, we obtain a similar simulation model whose outcome is in the same regime. The present simulation thus ensures that the new mass estimates remain quantitatively consistent with the scenario of a merger-induced dissociation.

\begin{table}
\caption{Initial condition parameters of the hydrodynamical simulation.}
\label{tb:ic}
\begin{center}
\begin{tabular}{l c c c c c}
\hline
\hline
           & $r_{h}$ & $r_{c}$ & $c$ & $M_{200}$            & $r_{200}$\\ 
           & (kpc)   & (kpc)   &     & ($M_{\odot}$)        & (kpc)     \\
\hline
A1758NW & 590     & 340     & 3.9 & $7.8 \times 10^{14}$ & 1844     \\
A1758NE & 500     & 230     & 4.2 & $5.4 \times 10^{14}$ & 1629     \\  
\hline
\hline
\end{tabular}
\end{center}
\end{table}


\section{The nature of the merging system A1758N}
\label{sec:discussion}

\subsection{Mass distribution}

Our comprehensive weak lensing analysis has enabled us to constrain well the total mass of the  Northern structure of A1758 (NW+NE) with a relative small error bar ($\sim10\%$ at 68 c.l.).  However, the individual mass measurements  (see Fig. \ref{fig:sub_mass}) point out that NW is more massive than NE within a probability  of $\sim79\%$.  This mass asymmetry helps explaining the feature observed, as the most massive structure was able to retain its gas whereas the subcluster was not. However, as pointed by \cite{Machado+2015}, some mass asymmetry is not required to explain this scenario, as it can be caused by original differences in the gas concentration, within some expected scatter.

We also found out that the positions of both Northern BCGs coincides with those of the respective lensing inferred mass centres within $2\sigma$.

On a more technical note, we found out that the use of the magnification bias with the distortion analysis helped decreasing the uncertainties on the masses to $\sim 80\%$ of the distortion only value.  On the other hand, the magnification effect showed very little power to constrain the position of the mass centres.

\subsection{Kinematics}

Examining the velocity distribution of the A1758N field (Fig. \ref{fig:vdistall}) we noticed two apparent galaxy concentrations, apart from the main one: one with 13 galaxies around $\bar{z}_{(1)}=0.328$ with $\sigma_{(1)}/(1+z)=326$ km s$^{-1}$ and other with 12 galaxies and $\bar{z}_{(2)}=0.377$  with $\sigma_{(2)}/(1+z)=518$ km s$^{-1}$. Both sets are very significantly away from A1758N and can be considered as separate entities. Since background massive structures could add some spurious mass in our mass reconstruction of A1758N, we estimated an eventual contribution of these structures. In a self similar scenario, the mass in a virialized cluster is proportional to $\sigma^3$ \citep{calberg97}; using our data, we found that these structures would contribute respectively with only $\sim 1\%$ and $\sim 5\%$ of the mass estimation of A1758N, well within the noise.

Regarding the velocity separation of the two components, the normality of the initial sample was a hint that it would be a very difficult task, as it could mean an absence or at least a small separation in the line-of-sight velocity between the two collisional structures. In fact, \cite{pinkney} had argued that one-dimensional  tests are more sensitive to identify substructures in systems where the merging is taking place close to the line-of-sight. Therefore, our lack of success to disentangle the main structures in redshift space may be seen as an evidence that the merging between A1758NW \& A1758NE is happening close to the plane of the sky. 

In this scenario the inclusion of the galaxy projected positions should be of help. However, the spectroscopic sample in not unbiased in the projected space. The mask making procedure forces a more homogeneous distribution than the underlying one, as it is harder to assign slits for objects too close to each other. For this reason we opted to use photometric selected red cluster galaxies. The separation of the two clumps in 2-D space yielded a line-of-sight relative velocity $\delta v/(1+z)=423\pm275$ km s$^{-1}$ between the two Northern components.  Together with our estimation of the relative velocity between the dark matter clumps, $v_{DM}=1100\pm100$ km s$^{-1}$,  we can estimate the angle $\theta$ between the merger axis and the plane of the sky as:
\begin{equation}
\theta=\arctan \left( \frac{\delta v}{v_{\rm DM}} \right) = 21\pm12\, \text{degrees}.
\label{eq:angle}
\end{equation}
%
To understand the effect of the merger on the dynamics of those systems we can compare the pre and post-collision velocity dispersions.
The latter is directly measured, whereas the former can be estimated by the weak lensing masses,  assuming mass conservation since the pre-collision up to the observation. For that we have used the proper scaling relations by \cite{biviano06} and the full mass posteriors (Fig. \ref{fig:mass_posterior}). We thus obtained a line-of-sight velocity dispersion ($\sigma_{line-of-sight}=\sigma_{3D}/\sqrt{3}$) of $1031_{-74}^{+80}$ km s$^{-1}$ and $912_{-82}^{+89}$ km s$^{-1}$ (for NW \& NE respectively), which leads to a boost factor, $f\equiv\sigma_{obs}/\sigma_{pre}$, of $f_{NW}=1.26^{+0.18}_{-0.19}$ and $f_{NE}=1.18^{+0.23}_{-0.20}$. Both values favour a young post-merger scenario \citep[Fig. 19 of][]{pinkney}. One also must note that a no boost scenario cannot be discarded within  $95\%$ c. l., considering all the uncertainties.

\subsubsection{Comparison with the previous studies}
\label{sec:comparison}

The previous weak lensing \citep{ragozzine} and dynamical \citep{boschin} analysis have not classified the member galaxies according to their host subcluster. In this paper, we have determined the individual properties (mass and velocity dispersion) of each structure of the merging system A1758NW \& A1758NE. As a result, we could compare quantitatively the gas, dark matter and galaxies distribution (traced by BCG position) in addition to constraining the inputs for a new hydrodynamical simulation.

Overall, the observed galaxy, intra-cluster gas and dark matter distributions are qualitatively consistent among the several studies. The identified red sequence galaxies (e.g. Fig.~\ref {fig:2d1d}) shows a bimodal behaviour with the clumps almost centred on the BGCs  in agreement with previous studies \citep{ragozzine,boschin,harvey15}. According to the mass reconstruction presented by \cite{ragozzine} the relative position in relation to the X-ray component are qualitatively coherent with our results.

Regarding the identification of the peak of the A1758NE X-Ray counterpart we are in disagreement with \cite{harvey15} which leads to very different estimates of the gas detachment. Whereas we and other authors \citep{sevenmergers,ragozzine,boschin,Machado+2015} identified the peak as in Fig.~\ref{fig:result}, Harvey and collaborators choose a local maximum of the X-ray emission substantially closer to the local BCG. By checking the  Chandra image we consider this second choice less likely given that the overall excess emitting region is smaller than cluster-scale and could possibly be a barely resolved structure superimposed to the diffuse cluster emission.

However, we notice that the peak of the X-ray clump related to A1758NE might have been misidentified in \cite{harvey15}, where it seems like a point source somewhat farther from the previously identified A1758NE X-Ray counterpart (e.g.~Fig.~\ref{fig:result}).

\subsection{The dark matter self-interaction cross section}

From our weak-lensing model, we have estimated a mean surface density $\overline{\Sigma}_s = 0.172$ g cm$^{-2}$ inside a radius of $R\leq 150$ kpc\footnote{This radius is the same used by \citet{markevitch04,merten,babybullet}.} in relation to the A1758NE mass centre. Using this value in Eq. \ref{eq:cross_section} it leads to an upper limit of the dark matter self-interaction cross section of 5.83 cm$^2$ g$^{-1}$. This limit is similar to the one found with the ``Bullet Cluster'' \citep[$\sigma /m<5$ cm$^2$ g$^{-1}$, ][]{markevitch04}, the ``Baby Bullet'' \citep[$\sigma /m<4$ cm$^2$ g$^{-1}$; ][]{babybullet} and Abell~2744 \citep[$\sigma /m<3 \pm 1$ cm$^2$ g$^{-1}$, ][]{merten}. 

Given our analysis, we found that the BCGs and dark matter centre positions are comparable within 95\% c.l. in the sense that it is not possible to extract trustworthy constraints to $\sigma/m$ through  the methodology presented by \cite{harvey15}.

\section{Summary}
\label{sec:conclusion}

The main findings of this work can be summarized as follows:

\begin{itemize}
\item[--] By modelling the weak gravitational lensing signals coming from distortion and magnification we were able to constrain the total mass of A1758N as $1.3^{+0.1}_{-1.4}\times 10^{15}$ M$_\odot$.

\item[--] The two sub-structures of A1758N have comparable masses, but there is a probability of 79\% that A1758NW is more massive than A1758NE. We measured the masses  $7.90_{-1.55}^{+1.89}$ $\times 10^{14}$ M$_{\odot}$  and $5.49_{-1.33}^{+1.67}$ $\times 10^{14}$ M$_{\odot}$, respectively for A1758NW and A1758NE.

\item[--] The magnification signal helped decreasing the uncertainties on the masses by $\sim20$\%, but did not help constraining the central positions of the sub-structures.

\item[--] In A1758NW there is a spacial coincidence between the positions of the local BCG, the peak of the X-ray emission and the centre of the lensing modelled mass clump centre. For A1758NE the BCG and mass centre positions are in agreement within 2$\sigma$ and both are very significantly away from the local maximum of the X-ray emission.

\item[--] The radial velocities of A1758N have shown a Gaussian distribution. However, giving this data, we cannot discard the hypothesis of a bimodal system with a small line-of-sight separation, which is a likely possibility if the merger is taking place near the plane of the sky, as suggested by previous X-rays analysis \citep{davidkempner}.

\item[--] With the space projected red sequence galaxies distribution we could find a bimodal scenario, with each structure related to A1758NW and A1758NE; they have, respectively, $\sigma_{\rm NW}=1296$ km s$^{-1}$ and $\sigma_{\rm NE}=1075$ km s$^{-1}$ with a line-of-sight separation of $\delta v / (1+z)=423\pm 275$ km s$^{-1}$.

\item[--]  The plane-of-the-sky relative velocity between the dark matter clumps, $v_{\rm DM}=1100\pm 100$ km s$^{-1}$,  based on our numerical simulation,  allowed us to estimate that the merging is taking place in just $21\pm12$ degrees from the plane of the sky. Moreover, the simulation pointed out that A1758N is young merger and it has been seen $0.27$ Gyr after the point of the closest approach of a  non head-on collision \citep{Machado+2015}.

\item[--] The 27 available radial velocities of A1758S show a unimodal distribution. On the other hand, due to the small number of galaxies the bimodality cannot be discarded. Searching for two spatially projected overlapped structures (as was pointed by previous \cite{davidkempner} X-rays analysis) we have found a scenario compatible with a merging happening very close to the line-of-sight.

\end{itemize}

\section*{Acknowledgements}

The authors thank Cristiano Sousa and Sarah Bridle for their collaboration in earlier phases of this project,  and Maru\v{s}a Brada\v{c}, Julian Merten and Joe Zuntz for valuable advices. RMO thanks the finantial support provided by CAPES and CNPq (142219/2013-4) and the project \textit{Casadinho} PROCAD - CNPq/CAPES (552236/2011-0). The authors would also like to acknowledge support from the Brazilian agencies CNPq and FAPESP (\textit{projeto temático} PI:LSJ 12/00800-4; ESC 2014/13723-3; REGM 2010/12277-9; ALBR 309255/2013-9). REGM also acknowledges support from \textit{Ci\^encia sem Fronteiras} (CNPq). 

This paper is based in part on observations obtained at the Gemini Observatory, which is operated by the Association of Universities for Research in Astronomy, Inc., under a cooperative agreement with the NSF on behalf of the Gemini partnership: the National Science Foundation (United States), the National Research Council (Canada), CONICYT (Chile), Ministerio de Ciencia, Tecnolog\'{i}a e Innovaci\'{o}n Productiva (Argentina), and Minist\'{e}rio da Ci\^{e}ncia, Tecnologia e Inova\c{c}\~{a}o (Brazil), observing run ID: GN-2010A-Q-22. Based in part on data collected at Subaru Telescope, via the time exchange program between Subaru and the Gemini Observatory. The Subaru Telescope is operated by the National Astronomical Observatory of Japan, observing run ID: GN-2007A-C-21.

This work has made use of the computing facilities of the Laboratory of Astroinformatics (IAG/USP, NAT/Unicsul), whose purchase was made possible by the Brazilian agency FAPESP (grant 2009/54006-4) and the INCT-A.

We made use of the NASA/ IPAC Extragalactic Database, which is operated by the Jet Propulsion Laboratory, California Institute of Technology, under contract with NASA.

\bibliographystyle{mn2e.bst}
\bibliography{mnras.bib}
\bsp

\onecolumn

\appendix
\begin{center}

\section{A1758N catalogue}
\label{ap:A1758Ncatalogue}

  \begin{longtable}{@{}cccccccccc@{}}
    \caption[]{Catalogue of the heliocentric radial velocities in A1758N field. In the first three columns we have, respectively, the galaxy ID, the galaxy membership ID and the ID adopted by \cite{boschin}, in order to ease the comparison for common objects. $B$, $R_C$ and $z'$ magnitudes are shown in next three columns and the radial velocity and its uncertainties are in the two last columns, in units of km s$^{-1}$. The galaxies marked with $^\star$ and $^\diamondsuit$ are the A1758NW e A1758NE BGCs respectively.}\\
    \hline \hline
	ID & IDm & IDb &$\alpha$ (J2000)& $\delta$ (J2000) & $B$ & $R_C$. & $z'$ & $v$ & $\sigma_v$ \tabularnewline    \hline
    \endfirsthead
    
    \multicolumn{10}{c}%
    {{\bfseries \tablename\ \thetable{} -- continued from previous page}} \tabularnewline
    \hline
     ID & IDm & IDb &$\alpha$ (J2000)& $\delta$ (J2000)& $B$ & $R_C$. & $z'$ & $v$ & $\sigma_v$ \tabularnewline
   \hline
    \endhead
    
    \multicolumn{10}{c}{{Continued on next page}} \tabularnewline \hline
    \endfoot  
    
   \hline \hline
    \endlastfoot
    
            1.&        -&           -&13:32:51.9& +50:28:02.2&       22.23&       21.00&       20.56&     210994.&        330.\\
            2.&          1.&        -&13:32:53.4& +50:28:45.1&       21.64&       19.58&       19.00&      81008.&         56.\\
            3.&          2.&        -&13:33:06.1& +50:29:16.4&       23.12&       21.19&       20.71&      83395.&        329.\\
           4.&          3.&        -&13:33:03.5& +50:29:08.0&       23.02&       20.97&       20.41&      83654.&        119.\\
            5.&          4.&        -&13:32:56.7& +50:29:25.3&       23.03&       20.81&       20.16&      83449.&        189.\\
            6.&          5.&        -&13:33:02.9& +50:29:22.9&       22.47&       20.40&       19.84&      81707.&        191.\\
            7.&          6.&        -&13:32:46.6& +50:29:27.2&       21.30&       19.33&       18.72&      81478.&         81.\\
            8.&          7.&        -&13:32:47.1& +50:29:40.1&       21.78&       19.55&       18.91&      83598.&         97.\\
            9.&          8.&        -&13:32:39.8& +50:30:20.9&       23.21&       21.12&       20.62&      83617.&        264.\\
           10.&          9.&        -&13:32:52.8& +50:30:26.6&       20.72&       18.49&       17.82&      84154.&         74.\\
          11.&        -&        -&13:32:36.8& +50:30:33.0&       21.00&       18.94&       18.53&     112353.&        306.\\
           12.&         10.&        -&13:33:10.7& +50:30:22.7&       21.36&       19.21&       18.54&      82001.&         97.\\
           13.&        -&        -&13:33:11.0& +50:31:05.0&       23.01&       21.03&       20.40&      69930.&        101.\\
           14.&         11.&        -&13:32:52.5& +50:31:29.1&       21.74&       19.43&       18.85&      82615.&         77.\\
           15.&         12.&        -&13:32:50.3& +50:31:26.2&       22.27&       19.96&       19.27&      85382.&         63.\\
           16.&        -&        -&13:32:50.2& +50:30:53.0&       21.96&       20.03&       19.43&      34122.&         23.\\
           17.&         13.&        -&13:32:50.9& +50:30:57.2&       22.74&       20.73&       20.12&      83389.&        112.\\
           18.&         14.&        -&13:32:45.1& +50:31:39.6&       23.17&       21.00&       20.32&      83501.&        135.\\
          19.&         15.&        -&13:32:58.1& +50:31:35.5&       22.87&       20.98&       20.31&      84887.&         99.\\
           20.&        -&        -&13:32:51.4& +50:31:43.9&       22.54&       20.08&       19.64&      34191.&         23.\\
           21.&         16.&        -&13:33:07.4& +50:32:10.2&       22.31&       20.31&       19.95&      82385.&        179.\\
           22.&         17.&        -&13:32:51.9& +50:32:18.7&       22.11&       19.96&       19.35&      81904.&         60.\\
           23.&         18.&        -&13:33:08.4& +50:32:25.4&       22.82&       20.93&       20.52&      82113.&        330.\\
          24.&         19.&        -&13:32:46.0& +50:32:34.0&       21.68&       19.75&       19.15&      82500.&         85.\\
           25.&         20.&        -&13:32:57.7& +50:32:36.5&       23.11&       21.05&       20.34&      83420.&        136.\\
           26.&         21.&        -&13:32:54.4& +50:33:34.2&       21.83&       19.96&       19.29&      81817.&        108.\\
           27.&         22.&        -&13:32:56.0& +50:32:49.1&       20.57&       18.39&       17.69&      79387.&         69.\\
           28.&        -&        -&13:32:25.8& +50:32:14.2&       22.57&       20.19&       19.45&     113112.&        330.\\
           29.&         23.&        -&13:32:39.6& +50:32:30.1&       21.78&       19.99&       19.40&      84052.&         73.\\
           30.&         24.&        -&13:32:38.9& +50:32:30.7&       21.98&       20.91&       20.69&      84633.&        118.\\
           31.&         25.&        -&13:32:43.2& +50:32:47.0&       23.27&       21.54&       20.92&      83537.&        202.\\
           32.&        -&        -&13:32:47.6& +50:32:46.4&       21.13&       20.44&       20.15&      45508.&         67.\\
           33.&         26.&        -&13:32:27.4& +50:32:36.9&       21.98&       19.72&       19.00&      83210.&        118.\\
           34.&         27.&        -&13:32:50.2& +50:33:21.4&       22.23&       21.61&       21.39&      82346.&         45.\\
          35.&         28.&        -&13:32:32.4& +50:33:33.8&       23.06&       20.91&       20.20&      85763.&         88.\\
           36.&         29.&        -&13:32:46.8& +50:33:18.3&       22.74&       20.72&       20.06&      87006.&         78.\\
           37.&        -&        -&13:32:39.4& +50:33:29.7&       22.73&       20.50&       20.52&      20868.&        119.\\
           38.&         30.&        -&13:32:37.0& +50:33:33.2&       22.15&       20.25&       19.71&      81647.&         69.\\
          39.&         31.&        -&13:32:32.4& +50:33:49.1&       21.60&       19.44&       18.69&      83757.&         97.\\
           40.&        -&        -&13:32:30.8& +50:33:51.1&       22.17&       20.65&       20.18&     169659.&        316.\\
           41.&        -&        -&13:32:15.0& +50:33:55.9&       22.12&       20.74&       20.20&      30509.&         96.\\
           42.&         32.&        -&13:32:43.6& +50:33:45.3&       22.14&       20.26&       19.69&      87050.&         77.\\
           43.&         33.&        -&13:32:39.6& +50:34:00.3&       21.65&       19.38&       18.63&      84995.&        100.\\
          44.&         34.&        -&13:32:45.0& +50:34:05.9&       21.79&       19.56&       18.82&      86582.&         63.\\
           45.&         35.&        -&13:32:49.0& +50:34:09.3&       21.65&       19.61&       18.94&      82313.&         53.\\
           46.&        -&        -&13:32:27.4& +50:34:03.7&       21.48&       19.29&       18.57&     113537.&         96.\\
           47.&        -&        -&13:32:41.1& +50:34:44.7&       22.24&       20.62&       19.93&     113510.&         96.\\
           48.&        -&        -&13:32:36.9& +50:34:52.4&       23.26&       21.69&       21.06&      97987.&         68.\\
          49.&         36.&        -&13:32:46.2& +50:35:08.9&       22.15&       20.02&       19.29&      85096.&         65.\\
           50.&        -&        -&13:32:34.8& +50:35:30.3&       22.81&       21.21&       20.63&      24198.&        121.\\
           51.&         37.&        -&13:32:38.1& +50:35:19.5&       22.17&       20.20&       19.50&      84309.&         53.\\
           52.&         38.&        -&13:32:46.8& +50:35:28.3&       22.56&       20.51&       19.81&      85339.&         76.\\
          53.&        -&        -&13:32:33.3& +50:35:29.8&       21.63&       20.65&       20.54&      74749.&         92.\\
           54.&        -&        -&13:32:34.9& +50:35:17.2&       21.20&       18.99&       18.25&      98316.&         76.\\
           55.&         39.&        -&13:32:35.9& +50:35:26.4&       22.34&       20.44&       19.70&      83210.&        115.\\
           56.&        -&        -&13:32:34.0& +50:35:42.8&       22.72&       21.18&       20.56&     113711.&        181.\\
           57.&        -&        -&13:32:33.6& +50:35:52.4&       23.10&       21.52&       21.01&       4210.&        175.\\
          58.&         40.&        -&13:32:35.0& +50:35:49.5&       21.51&       19.95&       19.38&      87509.&        369.\\
           59.&        -&        -&13:32:44.7& +50:36:04.5&       20.58&       19.90&       19.66&      53333.&        156.\\
           60.&        -&        -&13:32:41.6& +50:36:08.3&       20.96&       19.79&       19.56&      98092.&         60.\\
           61.&        -&        -&13:32:38.2& +50:36:32.9&       22.08&       21.04&       20.68&      17764.&        113.\\
           62.&         41.&        -&13:32:36.4& +50:36:50.4&       23.12&       21.54&       20.90&      86859.&        314.\\
           63.&         42.&        -&13:32:32.4& +50:37:04.9&       23.17&       21.41&       20.77&      86401.&        100.\\
           64.&         43.&        -&13:32:36.7& +50:37:08.6&       22.76&       21.59&       21.23&      86542.&         30.\\
           65.&        -&        -&13:33:00.3& +50:27:41.1&       22.33&       20.72&       19.84&      34410.&        176.\\
           66.&        -&        -&13:33:13.6& +50:31:11.1&       19.79&       17.72&       17.16&      70082.&        100.\\
           67.&        -&        -&13:33:11.3& +50:31:44.2&       22.44&       21.00&       20.46&      69950.&         65.\\
           68.&        -&        -&13:32:59.9& +50:33:44.5&       21.49&       19.77&       19.04&      49680.&        127.\\
           69.&        -&          2.& 13:32:14.6& +50:34:52.3&       21.04&       19.64&       19.06&      98743.&        590.\\
           70.&        -&         11.& 13:32:19.4& +50:34:11.2&       22.47&       20.16&       19.37&     112302.&        330.\\
          71.&        -&         15.& 13:32:22.5& +50:35:27.4&       21.94&       19.76&       19.10&      98439.&        109.\\
           72.&         44.&         16.& 13:32:22.8& +50:36:04.8&       22.24&       20.16&       19.57&      85119.&         74.\\
           73.&         45.&         17.& 13:32:25.2& +50:34:03.6&       20.49&       18.29&       17.62&      82854.&         79.\\
          74.&         46.&         21.& 13:32:29.2& +50:34:10.1&       21.42&       19.46&       18.88&      83384.&        105.\\
           75.&        -&         24.& 13:32:30.2& +50:36:37.3&       22.61&       20.24&       19.51&     113605.&         61.\\
           76.&         47.&         25.& 13:32:30.2& +50:33:49.2&       21.96&       19.74&       19.06&      83042.&        118.\\
           77.&        -&         27.& 13:32:31.6& +50:36:17.7&       21.54&       19.51&       18.79&      98488.&         53.\\
          78.&         48.&         30.& 13:32:32.8& +50:36:04.6&       21.98&       19.89&       19.17&      85182.&        110.\\
           79.&        -&         31.& 13:32:33.2& +50:34:02.6&       22.36&       21.26&       21.03&     187838.&       1025.\\
           80.&        -&         32.& 13:32:33.7& +50:31:03.6&       21.81&       19.79&       19.04&     112092.&        330.\\
           81.&         49.&         35.& 13:32:34.2& +50:33:03.2&       21.32&       19.26&       18.59&      87244.&         56.\\
           82.&         50.&         37.& 13:32:34.4& +50:33:18.3&       20.94&       18.78&       18.08&      84919.&         79.\\
           83.&         51.&         39.& 13:32:34.6& +50:31:39.2&       22.53&       20.25&       19.51&      84524.&        141.\\
           84.&         52.&         40.& 13:32:34.6& +50:33:24.8&       21.47&       19.22&       18.48&      81245.&         73.\\
           85.&        -&         46.& 13:32:35.7& +50:34:50.1&       22.82&       20.66&       19.89&     114055.&         97.\\
           86.&         53.&         48.& 13:32:36.4& +50:35:50.7&       21.47&       20.25&       19.73&      87377.&        108.\\
           87.&         54.&         51.& 13:32:37.5& +50:33:05.8&       21.41&       19.18&       18.41&      82414.&         73.\\
           88.&         55.&         53.& 13:32:38.3& +50:31:30.4&       22.60&       20.67&       20.11&      86371.&         63.\\
           89.$^\star$&  56.&         55.& 13:32:38.3& +50:33:35.8&       19.65&       17.41&       16.74&      83378.&         88.\\
           90.&         57.&         56.& 13:32:38.4& +50:31:41.5&       22.65&       20.59&       20.01&      85048.&        121.\\
           91.&         58.&         57.& 13:32:38.4& +50:32:53.2&       22.08&       19.99&       19.24&      83306.&        103.\\
           92.&         59.&         59.& 13:32:39.4& +50:34:45.2&       21.11&       18.89&       18.15&      83187.&         98.\\
           93.&         60.&         61.& 13:32:39.5& +50:34:32.1&       20.99&       18.84&       18.14&      87878.&         62.\\
           94.&         61.&         64.& 13:32:39.9& +50:32:24.0&       22.60&       20.57&       20.05&      83450.&        146.\\
           95.&         62.&         65.& 13:32:40.3& +50:34:49.7&       21.30&       19.34&       18.59&      86178.&         90.\\
           96.&         63.&         66.& 13:32:40.4& +50:35:39.7&       20.57&       18.41&       17.69&      81678.&         65.\\
           97.&         64.&         68.& 13:32:40.5& +50:33:15.5&       21.56&       19.39&       18.67&      80456.&        103.\\
           98.&         65.&         69.& 13:32:40.9& +50:33:46.4&       20.02&       17.78&       17.09&      83392.&         86.\\
           99.&         66.&         74.& 13:32:43.2& +50:29:26.1&       21.11&       18.85&       18.19&      82641.&         53.\\
          100.&         67.&         75.& 13:32:43.4& +50:33:28.8&       20.81&       18.52&       17.79&      85425.&         66.\\
          101.&         68.&         76.& 13:32:43.7& +50:31:14.7&       21.29&       19.23&       18.58&      83354.&        118.\\
          102.&         69.&         77.& 13:32:44.7& +50:33:26.1&       21.92&       19.68&       18.92&      84234.&         63.\\
          103.&         70.&         78.& 13:32:44.9& +50:31:57.5&       21.94&       19.82&       19.12&      84302.&         99.\\
          104.&         71.&         80.& 13:32:45.9& +50:32:04.8&       21.49&       19.30&       18.61&      79987.&         63.\\
          105.&         72.&         86.& 13:32:48.6& +50:31:54.9&       22.05&       19.73&       19.00&      82663.&         79.\\
          106.&         73.&         87.& 13:32:48.6& +50:31:21.8&       21.44&       19.18&       18.49&      83193.&        108.\\
          107.&         74.&         88.& 13:32:49.2& +50:31:31.2&       22.14&       19.90&       19.16&      82121.&         67.\\
          108.&         75.&         89.& 13:32:49.2& +50:28:50.4&       22.10&       20.66&       20.12&      82981.&        295.\\
          109.&         76.&         90.& 13:32:49.3& +50:33:56.1&       21.80&       19.91&       19.29&      80175.&         87.\\
          110.&         77.&         92.& 13:32:49.9& +50:32:26.0&       21.89&       19.89&       19.26&      84321.&        118.\\
          111.&         78.&         96.& 13:32:51.0& +50:33:08.8&       21.35&       19.31&       18.60&      84445.&         71.\\
          112.&         79.&         97.& 13:32:51.0& +50:27:46.0&       21.40&       19.54&       18.98&      81314.&         82.\\
          113.$^\diamondsuit$&         80.&        101.& 13:32:52.0& +50:31:33.9&       19.69&       17.36&       16.73&      83843.&         84.\\
          114.&         81.&        102.& 13:32:52.1& +50:31:22.0&       22.02&       19.99&       19.48&      83227.&        165.\\
          115.&         82.&        105.& 13:32:52.9& +50:31:46.1&       20.54&       18.26&       17.66&      79625.&         55.\\
          116.&         83.&        112.& 13:32:53.9& +50:32:20.7&       20.68&       19.00&       18.47&      79732.&         77.\\
          117.&         84.&        116.& 13:32:55.1& +50:31:26.1&       20.82&       18.60&       17.92&      85010.&         96.\\
          118.&         85.&        117.& 13:32:55.5& +50:31:28.5&       21.09&       18.84&       18.15&      85132.&        100.\\
          119.&         86.&        119.& 13:32:56.0& +50:30:17.5&       19.89&       18.88&       18.53&      82859.&         85.\\
          120.&         87.&        120.& 13:32:59.3& +50:30:20.0&       22.62&       20.55&       19.99&      83339.&        141.\\
          121.&         88.&        121.& 13:32:59.5& +50:29:27.6&       21.45&       19.25&       18.61&      81969.&         54.\\
          122.&         89.&        124.& 13:33:02.0& +50:29:28.0&       20.14&       17.89&       17.22&      83430.&         91.\\
          123.&         90.&        125.& 13:33:02.1& +50:30:09.3&       22.74&       20.67&       20.11&      83480.&        152.\\
          124.&         91.&        126.& 13:33:02.7& +50:31:42.3&       20.52&       18.90&       18.25&      80596.&         63.\\
          125.&         92.&        128.& 13:33:06.6& +50:31:24.3&       20.47&       18.50&       17.85&      85489.&        126.\\
          126.&         93.&        129.& 13:33:07.3& +50:29:19.6&       22.28&       20.15&       19.80&      82009.&        456.\\
          127.&         94.&        130.& 13:33:07.4& +50:32:00.7&       21.02&       18.94&       18.44&      82459.&        125.\\
          128.&         95.&        132.& 13:33:09.8& +50:29:02.3&       22.07&       20.12&       19.43&      83276.&        161.\\
          129.&         96.&        134.& 13:33:10.6& +50:31:15.7&       20.81&       18.76&       18.09&      86297.&        145.\\
          130.&         97.&        137.& 13:33:13.7& +50:29:55.2&       22.79&       20.68&       20.00&      83455.&        165.\\
          131.&        -&          1.& 13:32:13.9& +50:36:23.3&       23.01&       21.65&       21.05&     125775.&         71.\\
          132.&         98.&          3.& 13:32:15.3& +50:35:18.2&       21.92&       20.23&       19.54&      85168.&        104.\\
          133.&         99.&          4.& 13:32:16.8& +50:33:19.4&       21.41&       19.61&       19.05&      80993.&         94.\\
          134.&        -&          5.& 13:32:17.6& +50:37:27.2&       19.89&       18.99&       18.64&      41801.&        173.\\
          135.&        -&          6.& 13:32:17.8& +50:33:39.4&       20.86&       20.07&       19.74&      32788.&        100.\\
         136.&        100.&          7.& 13:32:17.9& +50:37:39.5&       21.62&       20.08&       19.48&      88373.&        128.\\
          137.&        -&          8.& 13:32:18.3& +50:36:47.6&       22.06&       20.10&       19.41&      98217.&         93.\\
         138.&        101.&          9.& 13:32:18.9& +50:36:04.4&       22.48&       21.01&       20.48&      84170.&        129.\\
         139.&        102.&         10.& 13:32:19.2& +50:36:03.9&       21.86&       19.96&       19.24&      85959.&        128.\\
          140.&        -&         12.& 13:32:22.2& +50:35:09.1&       21.63&       19.64&       18.93&      98001.&         50.\\
          141.&        -&         13.& 13:32:22.4& +50:34:48.8&       21.67&       19.82&       19.02&      97790.&         69.\\
          142.&        -&         14.& 13:32:22.4& +50:37:04.7&       23.51&       21.81&       21.11&      97872.&        100.\\
          143.&        103.&         18.& 13:32:25.9& +50:38:18.3&       21.70&       20.07&       19.46&      85091.&        114.\\
          144.&        104.&         19.& 13:32:27.5& +50:33:31.7&       21.96&       20.15&       19.48&      84801.&         98.\\
          145.&        105.&         20.& 13:32:27.5& +50:37:44.5&       21.64&       19.70&       19.08&      82907.&         92.\\
          146.&        -&         22.& 13:32:29.5& +50:35:26.0&       22.00&       20.46&       19.71&      74891.&        116.\\
          147.&        106.&         23.& 13:32:29.9& +50:37:21.6&       22.04&       19.86&       19.17&      82044.&         54.\\
          148.&        107.&         26.& 13:32:30.6& +50:36:25.5&       21.94&       19.91&       19.29&      82111.&         66.\\
          149.&        108.&         28.& 13:32:31.8& +50:34:50.0&       22.44&       20.17&       19.46&      83238.&        123.\\
          150.&        -&         29.& 13:32:32.6& +50:34:28.0&       20.31&       18.49&       17.86&      53251.&         53.\\
          151.&        -&         33.& 13:32:33.8& +50:30:50.1&       21.00&       19.01&       18.24&     112310.&         75.\\
          152.&        -&         34.& 13:32:34.0& +50:30:39.8&       21.87&       20.03&       19.45&     113091.&        116.\\
          153.&        109.&         36.& 13:32:34.3& +50:32:11.5&       21.18&       19.77&       19.60&      80234.&         41.\\
          154.&        110.&         41.& 13:32:34.7& +50:31:43.4&       21.96&       20.36&       19.81&      86327.&         79.\\
          155.&        111.&         42.& 13:32:34.9& +50:32:37.4&       20.66&       18.58&       17.91&      84194.&         44.\\
          156.&        -&         43.& 13:32:35.0& +50:38:30.3&       21.14&       19.61&       19.03&      98294.&        119.\\
         157.&        112.&         44.& 13:32:35.1& +50:32:36.4&       19.37&       17.96&       17.42&      81900.&         71.\\
          158.&        113.&         45.& 13:32:35.6& +50:33:38.2&       22.22&       20.03&       19.31&      83293.&         54.\\
         159.&        114.&         47.& 13:32:35.8& +50:33:52.1&       22.39&       20.36&       19.67&      83709.&         76.\\
          160.&        -&         49.& 13:32:36.8& +50:30:37.7&       22.21&       20.48&       20.06&     112681.&        100.\\
          161.&        115.&         50.& 13:32:36.8& +50:32:59.8&       21.48&       19.35&       18.60&      83963.&         63.\\
          162.&        116.&         52.& 13:32:38.0& +50:30:57.3&       22.64&       20.41&       19.78&      84344.&         79.\\
          163.&        117.&         54.& 13:32:38.3& +50:33:30.5&       20.88&       18.64&       18.04&      84176.&        120.\\
          164.&        118.&         58.& 13:32:38.5& +50:33:43.5&       20.72&       18.67&       18.08&      83791.&         79.\\
          165.&        119.&         60.& 13:32:39.5& +50:33:45.1&       22.00&       19.74&       19.16&      83880.&         76.\\
          166.&        120.&         62.& 13:32:39.7& +50:33:14.6&       22.36&       20.09&       19.37&      82872.&        128.\\
          167.&        121.&         63.& 13:32:39.7& +50:32:41.2&       21.05&       18.98&       18.30&      83803.&         84.\\
          168.&        122.&         67.& 13:32:40.5& +50:32:16.0&       23.67&       21.59&       20.93&      85932.&        182.\\
          169.&        123.&         70.& 13:32:41.4& +50:32:10.7&       21.82&       20.45&       20.15&      80544.&         57.\\
          170.&        124.&         71.& 13:32:42.0& +50:34:34.8&       19.62&       18.05&       18.16&      87429.&        132.\\
          171.&        125.&         72.& 13:32:42.1& +50:32:30.2&       22.48&       20.29&       19.54&      82939.&         85.\\
          172.&        126.&         73.& 13:32:42.3& +50:32:33.2&       22.68&       20.63&       19.96&      81072.&         91.\\
          173.&        127.&         79.& 13:32:45.3& +50:33:24.7&       21.28&       19.27&       18.56&      85798.&         82.\\
          174.&        128.&         81.& 13:32:46.3& +50:36:41.7&       20.89&       18.83&       18.12&      86035.&         47.\\
          175.&        129.&         82.& 13:32:46.8& +50:31:48.6&       22.07&       20.34&       19.61&      83825.&        113.\\
          176.&        130.&         83.& 13:32:46.9& +50:32:02.1&       21.62&       19.59&       18.94&      84249.&         76.\\
          177.&        131.&         84.& 13:32:47.9& +50:32:09.7&       21.88&       19.79&       19.15&      82909.&         82.\\
          178.&        -&         85.& 13:32:48.4& +50:28:06.7&       21.52&       19.21&       18.52&     109410.&         85.\\
          179.&        -&         91.& 13:32:49.7& +50:28:41.5&       20.06&       18.49&       17.89&      55954.&         60.\\
          180.&        132.&         93.& 13:32:50.0& +50:33:53.9&       21.37&       20.25&       19.87&      81273.&         60.\\
          181.&        133.&         94.& 13:32:50.5& +50:27:52.3&       20.85&       19.42&       18.93&      80403.&         72.\\
          182.&        134.&         95.& 13:32:50.7& +50:27:46.7&       21.37&       19.40&       18.83&      82691.&         79.\\
          183.&        135.&         98.& 13:32:51.4& +50:33:04.8&       19.47&       17.94&       17.41&      85536.&         57.\\
          184.&        -&         99.& 13:32:51.4& +50:33:10.4&       21.58&       19.62&       18.82&     130803.&        148.\\
          185.&        136.&        100.& 13:32:51.7& +50:27:13.6&       20.09&       17.89&       17.24&      83197.&         54.\\
          186.&        -&        103.& 13:32:52.3& +50:27:42.9&       22.43&       20.54&       19.87&     109756.&        204.\\
          187.&        137.&        104.& 13:32:52.3& +50:31:12.9&       22.30&       20.15&       19.50&      82717.&         75.\\
          188.&        138.&        106.& 13:32:53.0& +50:31:35.6&       21.09&       18.76&       18.17&      83648.&         63.\\
          189.&        -&        107.& 13:32:53.2& +50:30:29.0&       20.76&       18.48&       17.89&      99000.&         47.\\
          190.&        139.&        108.& 13:32:53.5& +50:27:36.0&       22.21&       20.33&       19.68&      83346.&         94.\\
          191.&        -&        109.& 13:32:53.6& +50:30:53.4&       20.48&       18.90&       18.32&      98877.&        104.\\
          192.&        -&        110.& 13:32:53.7& +50:26:56.2&       22.32&       20.78&       20.01&     212245.&        100.\\
          193.&        140.&        111.& 13:32:53.8& +50:29:57.8&       21.87&       19.75&       19.12&      82041.&         66.\\
          194.&        -&        113.& 13:32:54.7& +50:30:27.4&       21.57&       20.51&       20.39&      52848.&        100.\\
          195.&        141.&        114.& 13:32:54.9& +50:33:14.8&       21.86&       19.71&       19.00&      79373.&         54.\\
          196.&        -&        118.& 13:32:55.6& +50:30:55.5&       20.17&       18.67&       18.02&      55708.&         98.\\
          197.&        142.&        122.& 13:32:59.7& +50:30:05.6&       21.68&       20.60&       20.35&      86250.&        100.\\
          198.&        143.&        123.& 13:33:02.0& +50:27:17.1&       21.91&       19.73&       19.16&      82098.&         54.\\
          199.&        -&        127.& 13:33:03.6& +50:28:25.2&       21.41&       19.92&       19.35&      93943.&         91.\\
          200.&        144.&        131.& 13:33:09.7& +50:28:33.7&       20.84&       19.71&       19.34&      87007.&        100.\\
          201.&        145.&        133.& 13:33:10.1& +50:28:51.3&       20.99&       19.20&       18.67&      82496.&         72.\\
          202.&        146.&        135.& 13:33:10.8& +50:28:53.6&       21.75&       19.71&       19.03&      82857.&         91.\\
          203.&        147.&        136.& 13:33:12.9& +50:29:01.3&       20.51&       18.37&       17.68&      82798.&         41.\\
  \end{longtable}   
\end{center}

\clearpage

\begin{center}
\section{A1758S catalogue}
\label{ap:A1758Scatalogue}

  \begin{longtable}{@{}ccccccccc@{}}
    \caption[]{Velocity catalogue of A1758S.  First columns are, respectively, the galaxy ID and the galaxy member ID . $B$, $R_C$ and $z'$ magnitudes are shown in next three columns and the velocity and its uncertainty are in the two last columns  in units of km s$^{-1}$. The galaxy marked with $^\star$ is the Southern BGC.}\\
    \hline \hline
     ID & IDm &$\alpha$ (J2000)& $\delta$ (J2000) & $B$ & $R_C$. & $z'$ & $v$ & $\sigma_v$ \tabularnewline 
    \hline
    \endfirsthead
    
    \multicolumn{9}{c}%
    {{\bfseries \tablename\ \thetable{} -- continued from previous page}} \tabularnewline 
    \hline
     ID & IDm &$\alpha$ (J2000)& $\delta$ (J2000) & $B$ & $R_C$. & $z'$ & $v$ & $\sigma_v$ \tabularnewline 
   \hline
    \endhead
    
    \multicolumn{9}{c}{{Continued on next page}} \tabularnewline  \hline
    \endfoot  
    
   \hline \hline
    \endlastfoot

    1.&      1.       & 13:32:17.6& 50:22:17.7   &     22.25       &        19.86       &       19.16        &       81542.    &      58.  \\
    2.&      2.       & 13:32:33.5& 50:22:13.7   &     21.75       &        19.65       &       19.11        &       81112.    &      72.  \\
    3.&      3.		   & 13:32:26.5& 50:22:29.8   &     21.26       &        19.21       &        18.54       &       81540.    &      57.  \\
    4.&	    4.       & 13:32:34.6& 50:22:41.0   &     21.99       &        20.06       &        19.54       &       83557.    &     127.  \\
    5.&      5.       & 13:32:26.4& 50:22:57.9   &     23.49       &        21.54       &        20.90       &       83865.    &     297.  \\
    6.& 		6.	       & 13:32:26.0& 50:22:47.0   &    21.06       &        18.62       &        17.88       &       84009.     &     75.  \\
    7.&      7.       & 13:32:38.4& 50:23:09.6   &     22.38       &        19.87       &        19.14       &       83719.     &    106.  \\
    9.&     9.        & 13:32:18.9& 50:23:36.4   &     23.36       &        21.14       &        20.45       &       83078.     &    162.  \\
    10.&     -        & 13:32:12.8& 50:23:25.3   &     23.14       &        21.47       &        21.01       &       35622.     &    133.  \\
    11.&   10.      & 13:32:12.4& 50:23:43.1   &     22.38       &        20.00       &        19.25       &       83316.     &    125.  \\
    12.&     11.    & 13:32:12.5& 50:24:11.5   &     22.64       &        20.04       &        19.23       &       83278.     &    141.  \\
    13.&     12.    & 13:32:15.1& 50:24:06.3   &     20.67       &        18.61       &        17.96       &       84765.     &    121.  \\
    14.&     -        & 13:32:22.0& 50:23:58.2   &     20.83       &        18.51       &       17.84       &       69847.      &    52.  \\
    15.&     13.    & 13:32:33.2& 50:24:20.6   &     22.75       &        20.41       &        19.84       &       82326.     &     42.  \\
    16.&     -        & 13:32:34.0& 50:24:54.5   &     23.94       &        21.62       &        21.60       &       35446.     &     67.  \\
    17.&     14.    & 13:32:17.4& 50:24:38.6   &     21.60       &        19.42       &        18.76       &       80053.     &     73.  \\
    18.&     15.    & 13:32:31.4& 50:24:44.0   &     22.24       &        19.98       &        19.48       &       80716.     &     48.  \\
    19.&     16.    & 13:32:33.7& 50:24:49.9   &     21.90       &        19.93       &        19.46       &       82572.     &     78.  \\
    20.&     17.    & 13:32:26.3& 50:25:09.7   &     22.13       &        19.78       &        19.11       &       83064.     &     87.  \\
    21.&     18.    & 13:32:36.0& 50:25:23.4   &     22.83       &        20.77       &        20.17       &       81227.     &     50.  \\
    22.&     19.    & 13:32:21.4& 50:25:31.2   &     23.28       &        21.06       &        20.40       &       80550.     &     75.  \\
    23.&     20.    & 13:32:33.7& 50:25:36.3   &     22.93       &        21.14       &        20.67       &       79662.     &     78.  \\
    24.&     21.    & 13:32:24.5& 50:24:25.6   &     20.21       &        17.87       &       17.20       &       82655.      &    80.  \\
    25.&     -        & 13:32:36.5& 50:25:56.7   &     23.25       &        21.64       &        21.41       &       50151.     &    115.  \\
    26.&     -        & 13:32:37.7& 50:26:06.3   &     22.60       &        20.58       &        19.90       &      114113.    &     114.  \\
    27.&    22.     & 13:32:41.7& 50:25:15.0   &     21.31       &        19.25       &        18.69       &       82957.     &     50.  \\
    28.$ ^\star$&    23.     & 13:32:33.0& 50:25:02.4   &     19.52       &       17.21        &        16.76       &       81859.     &     59.  \\
    29.&    24.     & 13:32:14.4& 50:25:44.0   &     20.99       &        18.65       &       17.97       &       81205.      &    63.  \\
    30.&    25.     & 13:32:35.3& 50:26:23.9   &     21.55       &        19.58       &        19.01       &       80391.     &     42.  \\
    31.&    -         & 13:32:36.1& 50:26:46.1   &     22.97       &        20.76       &        20.07       &      114041.    &     227.  \\
    32.&    -         & 13:32:34.3& 50:26:51.1   &     21.91       &        20.66       &        20.28       &       55462.     &     81.  \\
    33.&     26.    & 13:32:28.8& 50:26:59.3   &     22.28       &        19.75       &        19.05       &       81727.     &     55.  \\
    34.&     -        & 13:32:41.0& 50:26:37.1   &     20.69       &        18.54       &        17.84       &       98944.     &     62.  \\
    35.&     27.    & 13:32:31.0& 50:27:14.0   &     20.45       &        18.60       &       18.15       &       81327.      &    43.  \\

  \end{longtable}   
\end{center}
\label{lastpage}

\end{document}